# "AI just keeps guessing": Using ARC Puzzles to Help Children Identify Reasoning Errors in Generative AI

Aayushi Dangol
University of Washington
Seattle, WA, USA
adango@uw.edu

Runhua Zhao[*]
University of Washington
Seattle, WA, USA
runhz@uw.edu

Robert Wolfe[*]
University of Washington
Seattle, WA, USA
rwolfe3@uw.edu

Trushaa Ramanan
University of Washington
Seattle, WA, USA
trushaar@uw.edu

Julie A. Kientz
University of Washington
Seattle, WA, USA
jkientz@uw.edu

Jason Yip
University of Washington
Seattle, WA, USA
jcyip@uw.edu

## Abstract

The integration of generative Artificial Intelligence (genAI) into everyday life raises questions about the competencies required to critically engage with these technologies. Unlike visual errors in genAI, textual mistakes are often harder to detect and require specific domain knowledge. Furthermore, AI's authoritative tone and structured responses can create an illusion of correctness, leading to overtrust, especially among children. To address this, we developed AI Puzzlers, an interactive system based on the Abstraction and Reasoning Corpus (ARC), to help children identify and analyze errors in genAI. Drawing on Mayer & Moreno's Cognitive Theory of Multimedia Learning, AI Puzzlers uses visual and verbal elements to reduce cognitive overload and support error detection. Based on two participatory design sessions with 21 children (ages 6 - 11), our findings provide both design insights and an empirical understanding of how children identify errors in genAI reasoning, develop strategies for navigating these errors, and evaluate AI outputs.

[*]denotes equal contribution

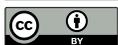

## 1 Introduction

As generative artificial intelligence (genAI) becomes increasingly integrated into educational environments, it presents both opportunities and challenges for teaching and learning [31, 61, 70, 89]. One revealing example comes from a middle school assignment on the novel *Persepolis*, where students researching prophets encountered flawed responses from internet searches that incorporated genAI. For instance, one response claimed that "*the Christian prophet Moses got chocolate stains out of T-shirts*" — a stark misunderstanding of historical and religious contexts [30]. According to the teacher, interviewed by the New York Times, more concerning than the error itself was that eighth graders accepted and recorded the AI hallucination without questioning its validity [30]. This incident underscores broader concerns in AI literacy research regarding children's trust in AI technologies and the critical need to equip them with the skills to engage effectively with AI, critically assess its outputs, and understand its strengths and limitations [41].

A key competency in this regard is understanding the difference between tasks that genAI performs well and those where it falters. While genAI excels at detecting patterns and generating fluent text, it struggles with applying knowledge in new contexts and reasoning through multi-step problems that require deeper understanding [94]. These limitations can lead to faulty reasoning and misleading outputs, which may reinforce misconceptions [5, 72, 85] and pose risks—ranging from misinformation in educational settings to flawed legal or medical recommendations [1, 3, 51, 71]. However, recognizing errors in genAI's outputs is not always straightforward [15, 69]. Even for adults, misleading outputs can be difficult to detect [74], and the challenge is often greater for children, who may have less experience questioning authoritative-seeming information [39, 68, 86].

One of the key challenges in detecting inconsistencies in AI-generated text is that, unlike images, textual responses do not present errors in an immediately perceivable visual pattern [54, 69]. In AI-generated images, inconsistencies like extra fingers or distorted facial features can be easily noticeable [8]. In contrast, recognizing textual inaccuracies often requires specific domain knowledge. Research by Solyst et al.[69] highlights that genAI systems like ChatGPT can create an illusion of correctness, even when their responses are inaccurate, by presenting neatly formatted outputs and seemingly logical explanations. These factors can lead middle school children to over trust AI-generated content [69]. Additionally, the length and verbosity of AI-generated responses can increase cognitive load, making it harder for children to identify inconsistencies [46, 54]. Furthermore, without explicit indicators of potential mistakes, they may struggle to assess the reliability of the information presented [4, 90]. Thus, prior work suggests that textual responses alone may not be an ideal starting point for children to detect genAI's limitations.

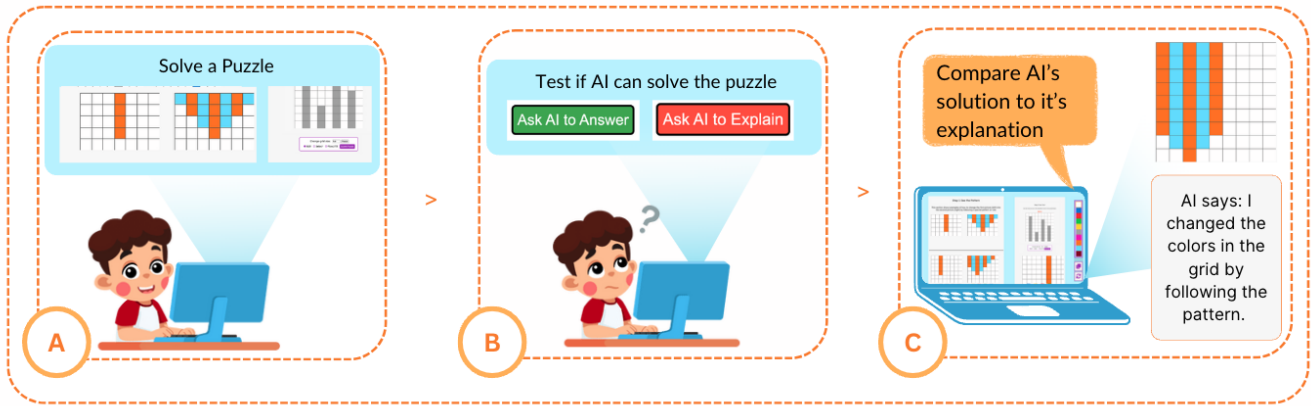

Figure 1: Overview of AI Puzzlers: (A) children first solve an ARC puzzle independently, then (B) test whether genAI can solve the same puzzle, and finally (C) compare AI's solution to its explanation to evaluate AI reasoning.

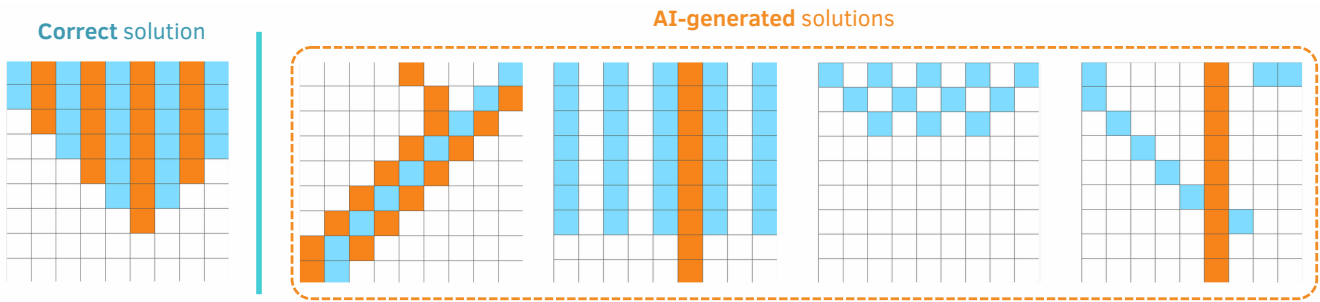

Figure 2: Comparison of the correct vs AI-generated solutions. The visual nature of AI Puzzlers makes AI errors easy to spot.

Recognizing that children are naturally drawn to games and puzzles [9, 12, 17, 82], we saw an opportunity to address this need by adapting Abstraction Reasoning Corpus (ARC) puzzles [14], originally developed to benchmark AI progress, into a web-based game called AI Puzzlers (see Figure 1). To design our system, we built on Mayer and Moreno's theory of multimedia learning, which emphasizes distributing information between visual and verbal channels to prevent cognitive overload [45, 46]. AI Puzzlers allows children to visually compare genAI's solution with their own and engage with AI-generated explanations (see Figure 2), providing opportunities to examine genAI's reasoning in relation to its visual output.

We then conducted two participatory design sessions with 21 children (ages 6–11) to answer the following research questions:

- What specific limitations of generative AI do children (ages 6–11) encounter and recognize through their engagement with AI Puzzlers?
- How does presenting information across visual and textual modalities influence children's ability to critically assess AI-generated outputs?

Our findings show that AI Puzzlers provided a tangible way for children to engage with genAI's reasoning by making its outputs visually comparable to their own. Even younger children, who were not yet fluent readers, quickly detected inconsistencies in AI-generated solutions by evaluating their visual outputs. When genAI made mistakes — especially on puzzles they considered easy — children reacted with surprise and amusement, sparking meaningful dialogue around how "*AI thinks*." This also helped them recognize that genAI approaches problem-solving differently from humans and, despite its strengths, has limitations that require careful evaluation of its outputs. Their continued engagement with AI Puzzlers highlights the importance of designing genAI systems that present information in ways that facilitate comparison, encourage reflection, and scaffold multiple ways of understanding. This way, children are more likely to persist and critically evaluate AI outputs.

Our contributions offer both design insights and an empirical understanding of how children make sense of genAI's reasoning,

navigate its errors, and develop strategies for evaluating its outputs. In the following sections, we first discuss related work on AI literacy and children's interactions with AI. We then describe the design of AI Puzzlers, outlining its system design and theoretical foundations. Finally, we present our results and discuss their implications for supporting AI literacy interventions.

## 2 Related Work

### 2.1 Children's Interactions with Generative AI

While empirical data on children's use of generative AI is limited, early surveys suggest that children and youth are engaging with generative AI at increasing rates, often surpassing adult adoption [11, 48]. Within HCI, research has explored the integration of generative AI to support creativity [42, 54, 64], storytelling [32, 96], and learning [13, 18, 21, 70, 89], while also examining biases that may affect children and teenagers [84, 85]. Prior studies suggest that children's (ages 8–13) ability to critically assess AI-generated content is heavily influenced by their prior knowledge and age [19]. For example, children who are well-versed in specific topics, such as Pokémon or Star Wars, can readily identify errors when ChatGPT provides inaccurate information or when a DALL-E-generated image depicts a character with six fingers instead of five [54]. Conversely, when faced with unfamiliar subject matter, children are more likely to overtrust AI-generated outputs [69].

Prior research also suggests that children's tendency to over trust genAI stems largely from how it presents information [52]. Aesthetic legitimacy plays a key role as AI-generated outputs often appear polished, with neatly formatted step-by-step instructions or structured lists, creating an illusion of correctness [68, 69]. For example, middle school students found ChatGPT's response to a technical question convincing because of its clear organization of the text, even though the content was inaccurate [69]. Similarly, perceived transparency, where AI appears to explain its reasoning, can mislead users [68, 69]. In one example, children trusted ChatGPT's breakdown of a multiplication problem because the steps appeared logical, even though the answer was incorrect [69]. Superficial indicators of correctness further compound this problem, as people often rely on details like citations, dates, or names to judge accuracy [68, 69]. For example, a fabricated list of scientific papers generated by AI was trusted by students because it included plausible sounding titles and dates [69]. From a cognitive load perspective, these challenges arise because children have limited working memory resources, making it difficult to critically assess AI-generated content without proper scaffolding [45, 46]. Generative AI interfaces, like ChatGPT, lack built-in validity indicators, increasing the cognitive burden on users who must independently verify information [68]. Without structured guidance, children may struggle to discern errors, particularly when AI presents information in a coherent yet misleading way. Given these challenges, there is a growing need to support children in critically engaging with generative AI while managing cognitive load.

### 2.2 Multimedia Learning and AI Literacy for Children

Multimedia instruction involves presenting words and images (static or dynamic) to support learning. Mayer and Moreno's Cognitive Theory of Multimedia Learning (CTML) posits that humans process information through verbal and visual channels [44]. By distributing information across these channels, multimedia learning reduces cognitive load and enables learners to construct complementary verbal and visual mental models and form connections between them [44, 46]. Additionally, prior research demonstrates that integrating words and images helps with comprehension and retention compared to presenting words alone [45, 58, 95].

Given that AI literacy involves critically evaluating AI outputs and collaborating effectively with AI [41, 76], applying multimedia learning in AI literacy platforms can help learners process and assess AI-generated content without cognitive overload. To support AI literacy, scholars have applied multimedia learning strategies in various educational platforms. For example, platforms like *PoseBlocks* [35], *danceON* [59] and *Google's Teachable Machine* [10] enable children to train machine learning models using images and sounds, providing a hands-on way to explore concepts like classification and training data. Furthermore, prior research suggests that open-ended platforms where children can experiment with AI models can improve students' understanding of AI and encourage meaningful discussions about its capabilities and limitations [22, 23, 26, 78].

Several AI literacy initiatives have also used games as an effective form of multimedia learning. For example, Ng et al., developed *TreasureIsland*, an online educational game to teach AI concepts and AI ethics [55]. Their study showed that the game was effective in improving students' motivation, self-efficacy, career interest, and understanding of AI. Similarly, inter-generational games have also been used to incorporate both technical and ethical AI knowledge, providing a family-centered approach to AI literacy [69]. These findings highlight the potential of game-based learning to engage diverse audiences while scaffolding complex AI concepts. In the next section, we further examine the role of games in learning and its implications for child-genAI interaction.

### 2.3 Learning through Games

Within HCI, researchers have highlighted the potential of games for creating an engaging, low-pressure environment that encourages exploration and skill development [12, 17, 82, 98]. Prior research shows that games provide an interactive space where learners can experiment, receive immediate feedback, and refine their understanding through trial and error [29, 55]. Well-designed games incorporate principles of effective teaching by scaffolding learning experiences, adapting to different learning speeds, and offering structured opportunities for practice [16, 28, 40]. These qualities make games particularly well-suited for educational contexts, where motivation and engagement are crucial for sustained learning. Prior research has shown that scaffolding within educational games can reduce cognitive overload for novice AI learners by guiding learners' attention, and structuring problem-solving processes [38, 97]. For example, visual indicators of progress and interactive elements within games can direct attention to key AI concepts, thereby minimizing split-attention effects and improving comprehension [73]. Modular learning structures further break down complex AI concepts into smaller, more manageable components, allowing learners to gradually build their understanding [37]. These mechanisms ensure that challenges remain within the learner's zone of proximal development [79], allowing them to engage meaningfully with the material without becoming overwhelmed.

Considering the growing role of genAI in children's learning, fostering their ability to critically engage with AI-generated reasoning becomes increasingly important. This raises a significant question: *How might games help children develop an awareness of genAI's reasoning processes and limitations?* More specifically, we examine how engaging children in solving puzzle games and comparing their solutions with AI-generated responses can create moments of critical evaluation, foster skills necessary for AI literacy, and reduce cognitive load. Guided by Mayer and Moreno's CTML [44] and prior work in AI literacy, the next section introduces AI Puzzlers, an interactive game designed to help children critically evaluate AI-generated outputs and develop a deeper understanding of genAI's strengths and limitations.

## 3 AI Puzzlers: System Design & Development

AI Puzzlers is an interactive system designed to help children (ages 6+) critically engage with genAI's reasoning by solving visual puzzles. Accessible through any web browser, AI Puzzlers requires no prior knowledge of AI or programming, making it an easy entry point for young learners. The system's codebase is publicly available at https://github.com/adango26/Puzzleland. We developed AI Puzzlers using the Abstraction and Reasoning Corpus (ARC) dataset [14], a collection of 800 publicly available visual puzzles originally designed to assess AI's reasoning capabilities. As illustrated in Figure 3, to solve an ARC Puzzle, a human player or AI agent follows the following steps:

- **Infer the Transformation Rule:** Each puzzle presents two or more input grids and their corresponding output grids. By observing the example input-output pairs, players try to discover the hidden rule that explains how the input changes to create the output.
- **Apply the Rule to New Inputs:** Once the transformation rule is inferred, players apply it to new, unseen input grids to generate the correct output.

Our design builds on ARC puzzles because 1) solving these puzzles requires no prior knowledge, creating a low barrier to entry; 2) AI systems struggle with solving ARC Puzzles while humans excel, aligning with Long's AI literacy competency on recognizing when to leverage AI versus human strengths [41]; 3) children are naturally drawn to puzzles and games, making them an engaging game-based medium that can scaffold learning; and 4) the visual nature of the puzzles ensures there is no obscurity in the way genAI presents information. By visually comparing genAI's solution with the correct solution that children can easily solve, they can spot when genAI makes mistakes, preventing children from being misled by polished yet incorrect answers. This process also encourages critical evaluation of genAI outputs, allowing children to recognize not only the limitations of genAI but also the unique strengths of human reasoning. We further elaborate on our design considerations and how they align with learning theories in the section below.

### 3.1 Design Considerations

*3.1.1 Facilitating Critical Evaluation through Visual Comparisons of genAI Outputs.* Mayer and Moreno's active processing assumption suggests that meaningful learning occurs when learners actively select, organize, and integrate new information with prior knowledge [44, 45]. AI Puzzlers facilitates this process by allowing children to visually compare their own puzzle solutions with those generated by AI. This direct visual comparison engages children in deeper cognitive processing, prompting them to notice differences, identify errors and reasoning gaps in the AI's process. Immediate feedback from the system further reinforces active processing by helping children quickly see what worked, what did not work, and why, facilitating critical evaluation of genAI's outputs.

*3.1.2 Reducing Cognitive Overload in Interpreting genAI Outputs.* According to Mayer and Moreno's dual-channel assumption [44, 46], people process information through two distinct channels: visual and verbal. Distributing cognitive effort across both channels reduces cognitive overload, leading to more efficient processing and deeper understanding. AI Puzzlers leverages this principle by presenting both visual and textual explanations of genAI's reasoning. When children use the "Ask AI to Explain" feature, they receive a visual representation of the AI's solution along with a step-by-step explanation of its reasoning. This dual presentation helps children compare the AI reasoning with its visual output, making it easier to spot discrepancies and evaluate the genAI solution.

*3.1.3 Scaffolding Learning through Easy-Switch Modalities.* AI Puzzlers scaffolds learning and reduces cognitive overload by ensuring that children build foundational knowledge before tackling more advanced tasks [44, 46]. It supports three different modalities to help children develop familiarity with the puzzle environment before they engage with more complex tasks. In Manual Mode, children learn basic puzzle-solving tools to solve the ARC puzzles, building foundational skills without the complexity of AI interactions. AI Mode introduces features like "Ask AI to Solve," and "Ask AI to Explain" allowing children to gradually engage with AI, while Assist Mode lets them actively guide the AI through puzzle-solving, encouraging experimentation and deeper understanding. These modes are designed for seamless switching, allowing children to easily transition between them.

*3.1.4 Fostering Exploration of genAI's capabilities through Active Debugging.* AI Puzzlers supports children in critically engaging with genAI's reasoning by recognizing its mistakes and providing hints to guide its responses. This fosters a form of participatory debugging [40, 75], where children take on an active role in evaluating genAI's logic and suggesting improvements. This approach is informed by Wittrock's generative learning theory [83], which emphasizes active meaning-making, as children generate hypotheses, test AI responses, and construct new insights through comparison and reflection. Through the process of debugging, children refine their understanding of genAI's capabilities and limitations, aligning with Wittrock's notion that learning emerges through the active construction of relationships among ideas [83].

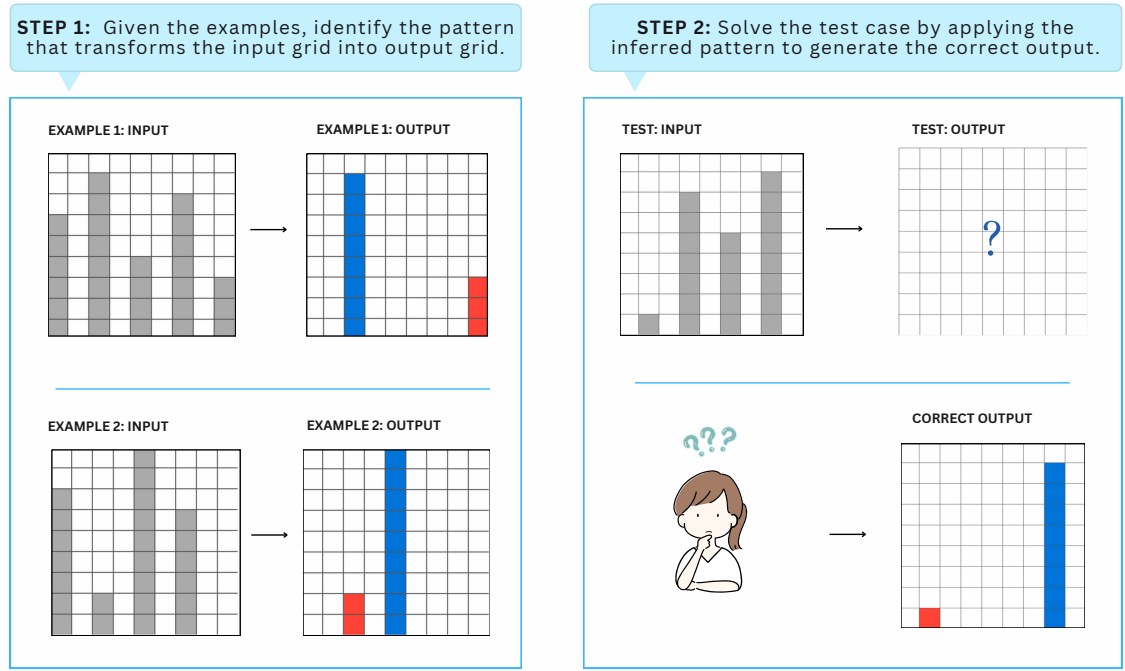

Figure 3: An example of an ARC puzzle with instructions for solving it. The correct answer shows the shortest bar is colored red and the tallest one is blue.

### 3.2 Design Overview

AI Puzzlers consists of 12 puzzles from the ARC dataset, distributed across four levels of difficulty. We play-tested them with $N = 106$ children (grades 3 - 8) during our university's annual K-12 STEM outreach event [20]. These children, who were event attendees (separate from our main study participants), voluntarily participated in the playtesting. After completing the puzzles, children rated their difficulty on a Likert scale from 1 (very easy) to 5 (very hard), with 3 as a neutral response. A one-sample $t$-test showed that children perceived the puzzles as slightly easier than neutral (M = 2.38, $t(103) = -6.48$, $p < .001$). To support children's critical engagement with genAI's outputs, AI Puzzlers employs a scaffolded interaction model with three modes, which we describe next.

*3.2.1 Manual Mode.* This is the default interaction mode upon launching the AI Puzzlers application (see Figure 4). It is designed to encourage children to engage with the ARC puzzles independently and provides a foundation that will be expanded upon in subsequent interaction modes. AI Puzzlers offers players several functionalities to solve the puzzle. First, players can customize the grid size by adjusting the row and column height of the "After" grid. Although the predefined "Before" grid is non-editable, children have the option to "clone" it into the "After" grid. A palette of 10 distinct colors is provided for use on the editable grid. The "Edit" tool allows players to modify the color of individual squares. The "Select" tool enables the highlighting of multiple squares simultaneously, facilitating batch modifications. The "Flood Fill" tool changes the color of all connected squares that share the same color, similar to the paint bucket tool commonly found in graphics programs. Additionally, a "Reset" button is available, enabling players to start over with a clean version of the "After" grid if they wish to try a different approach.

*3.2.2 AI Mode.* Building on the features of Manual mode, AI mode allows children to interact with GPT-4o through "Ask AI to Solve" and "Ask AI to Explain" features. By clicking the "Ask AI to Solve" button, players can have GPT-4o attempt to generate a solution, which is then displayed in the "After" grid. Under the hood, a `grid_parser()` function converts the visual puzzle into a textual grid representation that GPT-4o can interpret. The model then outputs a similar textual grid structure (e.g., [[0, 2, 2], [0, 1, 1], [0, 0, 3]]), which a `response_parser()` function translates back into a visual grid, using tokens for colors and positions. Because GPT-4o's responses are generated in real time via its API, the answers may vary with each attempt, demonstrating its generative nature. Additionally, the "Ask AI to Explain" feature provides step-by-step, child-friendly explanations of the AI's reasoning process, which can be toggled on or off according to the player's preference. After reviewing the AI's process, children can submit the AI's solution to receive immediate feedback on its correctness, just as they would with their own solutions. This process allows children to see where the AI might struggle, offering a direct and interactive way to understand the limitations of AI in solving ARC puzzles.

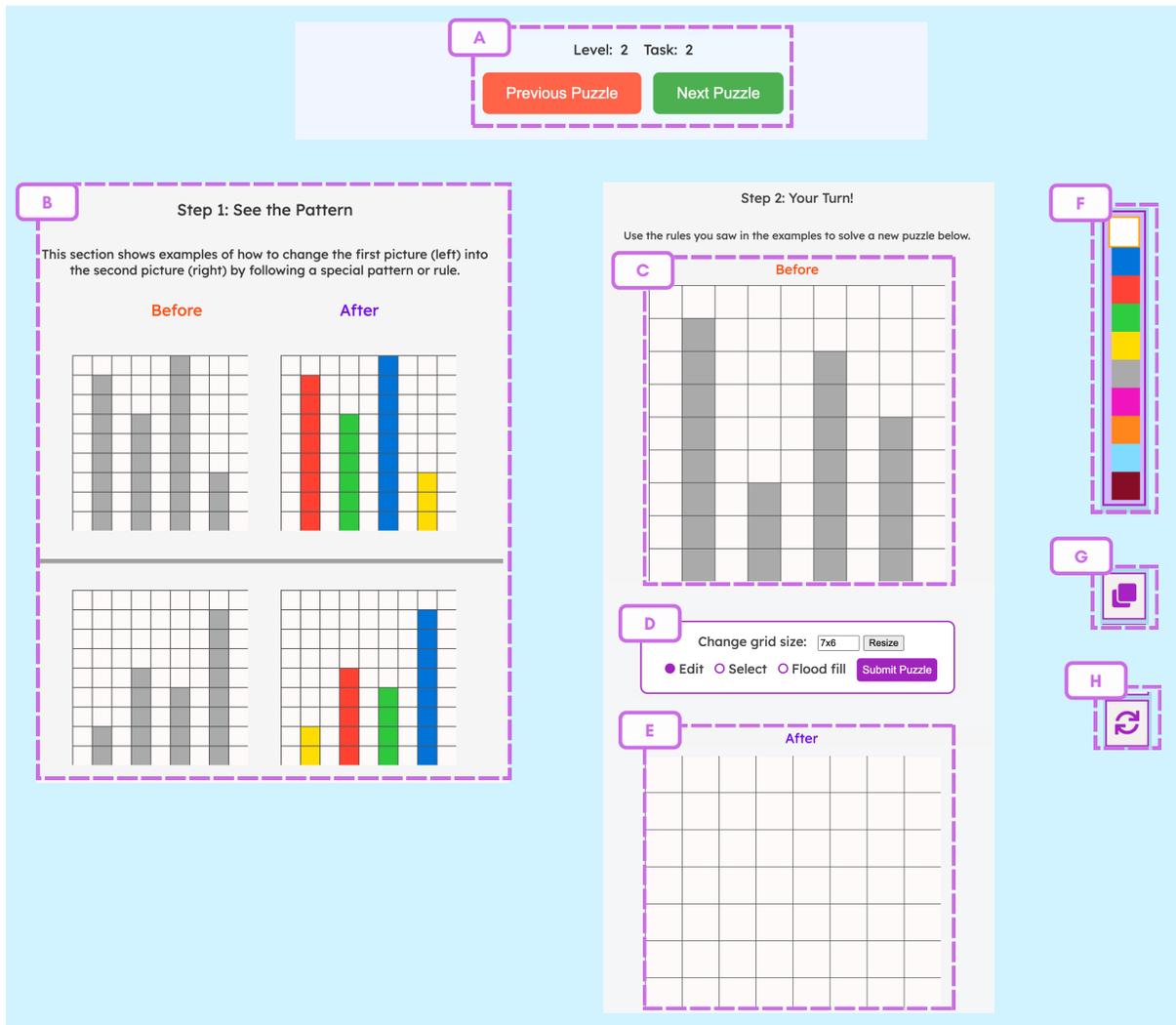

Figure 4: Annotated screenshot of Manual Mode in AI Puzzlers. The interface consists of several key components: (A) Puzzle Navigation allows children to switch between different puzzles. (B) Example Puzzles provide input-output pairs to illustrate transformation rules. (C) Before Grid presents the given input that children use to solve the corresponding After puzzle. (D) Edit Panel provides a set of tools for editing After grid. (E) After Grid serves as the solution space where children create their expected output based on the transformation rule. (F) Color Palette enables children to apply colors when solving the puzzle. (G) Clone Button transfers the Before Grid content to the After Grid for further modification. (H) Reset Button clears the After Grid to start over.

3.2.3 *Assist Mode.* Expanding further on the features of the AI mode, this mode allows children to guide the AI to solve puzzles by actively participating in the decision-making process and experimenting with different strategies to help the AI (see Figure 5). Similar to debugging in programming [40], children can identify mistakes in the AI's approach and suggest corrections by typing their suggestions to AI into a "Hint" field. For example, if the AI incorrectly sorts shapes, a child might type, "Look at the smallest and biggest shapes." The AI will then take these hints and produce a new output. This interactive process of testing and debugging helps children develop a more sophisticated understanding of AI's capabilities, as they see the direct impact of their input on the AI's performance. Children can also experiment with different puzzle configurations and observe how the AI responds to changes in input. They can adjust parameters such as: 1) alter the number of input-output examples provided to the AI, 2) test different versions of OpenAI's GPT models, including GPT-3.5-turbo, GPT-4o-Mini, GPT-4o, and GPT-4o-Turbo or 3) introduce random patterns without clear transformation rules.

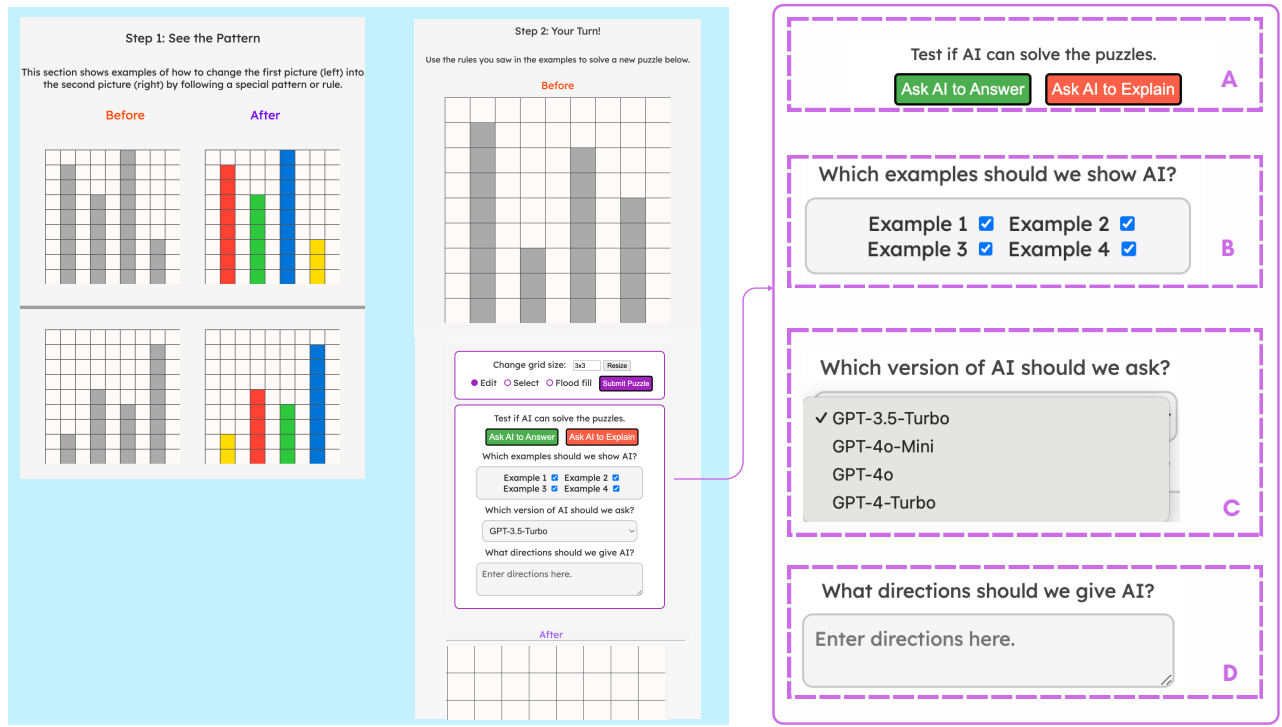

Figure 5: Screenshot of Assist Mode in AI Puzzlers. Certain interface elements are enlarged to highlight key interactive features, including (A) testing if AI can solve the puzzles by either answering or explaining, (B) selecting the number of examples shown to AI, (C) choosing AI model versions, and (D) providing hints to AI.

## 4 Methods

We employed a participatory design (PD) method called Cooperative Inquiry (CI) [24, 25, 93] to investigate how children used AI Puzzlers to critically engage with generative AI. Originating from Druin's work, CI builds on the principle that children possess unique expertise in being children and should be positioned as equal and equitable partners alongside adult researchers in the design of new technologies [24, 25]. As highlighted by Yip et al. [93], this equitable design partnership is grounded in four key dimensions: relationship building, which fosters trust and mutual respect; facilitation, which ensures children's ideas are valued and integrated; design by doing, which emphasizes hands-on iterative collaboration; and idea elaboration, where children refine and expand their contributions through active discussion and feedback.

We adopted CI as our methodological approach for several reasons. First, CI fosters a reflective and dialogic environment where children can vocalize their thought processes, negotiate perspectives, and collaboratively engage with AI Puzzlers. Second, CI has been widely applied in child-computer interaction research to examine how children conceptualize emerging technologies like intelligent interfaces and social robots [53, 56, 87, 88]. Prior research has shown that children who are comfortable working with adults can express their perceptions more assertively, allowing them to articulate abstract ideas in more concrete ways [92]. Finally, the co-design setting, where children are already familiar with multiple PD techniques, allowed us to observe in-the-moment decision-making quickly and efficiently [80]. This provided valuable insights into points of confusion, breakdowns in AI-generated explanations, and opportunities for refining system design in future work.

### 4.1 Participants

We conducted our study with an inter-generational co-design group called **KidsTeam UW** consisting of twenty-one children, ages 6 to 11 (M = 8.10, $\sigma$ = 1.45), and adult design researchers (researchers, graduate and undergraduate research assistants). Child participants represented a diverse range of ethnic backgrounds, including Asian, White, Black, Hispanic, and Multiracial identities. Children were recruited through mailing lists, posters, and snowball sampling. The child participants reported varying degrees of AI use and familiarity. While six children engaged with AI daily, three had no prior experience. The most common AI interactions included video game AIs and voice assistants. Table 1 presents the demographic information and AI usage details of the children, with all names represented as pseudonyms. Parental consent and child assent were obtained for all child participants, and the study was reviewed and approved by our university's Institutional Review Board.

### 4.2 Co-Design Sessions

We conducted two sessions with KidsTeam UW as part of a week long summer camp hosted by an inter-generational co-design group at our university. Each 1.5 hour session began with a 15-minute informal discussion to foster open dialogue and build rapport before transitioning to hands-on engagement with AI Puzzlers. Child participants worked in small, collaborative groups, each consisting of four to five children and two adult facilitators (see Figure 6). This structure was designed to balance peer-driven exploration with adult guidance, ensuring children could inspire one another while still receiving individualized support. To mitigate power imbalances and minimize undue influence on children's responses, adult facilitators were trained to promote equal participation, fostering a balanced dialogue where all children had the opportunity to contribute.

Table 1: Reported Child Participant Details

| Name | Gender | Ethnicity | Age | AI Type | Usage Frequency |
|---|---|---|---|---|---|
| Kai | Male | Asian/White | 8 | Voice Assistant | Daily |
| Lani | Female | Asian/Black | 9 | None | Never |
| Juno | Male | Asian | 7 | Video Game AIs, Voice Assistant | Daily |
| Elias | Male | Asian/Black | 9 | Video Game AIs, Voice Assistant | Daily |
| Noa | Female | Asian/White | 11 | Video Game AIs, Voice Assistant | Multiple times a week |
| Ren | Male | Hispanic | 10 | Chatbot | Multiple times a week |
| Matt | Male | Asian/White | 9 | N/A | N/A |
| Ivy | Female | White | 9 | Video Game AIs, Voice Assistant | A few times a week |
| Zayn | Male | Asian/Black | 9 | None | Rarely |
| Finn | Male | White | 10 | N/A | N/A |
| Leila | Female | Asian | 8 | Voice Assistant | Daily |
| Mara | Female | Asian/Black | 6 | Video Game AIs, Voice Assistant | A few times a week |
| Emi | Female | Asian/White | 8 | None | Rarely |
| Hana | Female | Asian | 8 | None | Multiple times a week |
| Theo | Male | Asian/White | 7 | Video Game AIs, Voice Assistant | Multiple times a week |
| Lucia | Female | Hispanic | 6 | Video Game AI | Weekly |
| Rina | Female | Asian | 7 | Video Game AIs, Voice Assistant | Monthly |
| Owen | Male | White | 8 | Video Game AI | Daily |
| Nico | Male | Asian/White | 6 | None | Daily |
| Selah | Female | Asian/Black | 6 | None | Never |
| Elise | Female | Asian/Black | 9 | None | Never |

4.2.1 *Session 1.* We began the session with a warm-up question: "*Tell us about someone or something that you think is smart—and why?*" This activity served as an icebreaker, allowing children to become familiar with the researchers and fostering an open discussion environment. Next, children were introduced to both the manual mode and AI mode of AI Puzzlers. The research team demonstrated how to solve ARC puzzles and explained that in AI mode, they could ask genAI to solve the puzzles. Earlier in the day, as part of the broader summer camp, children had already interacted with genAI by requesting and observing AI-generated images using *DALL·E 3*. Given this prior exposure, we chose to build upon their understanding rather than reintroduce genAI concepts. This allowed our session to focus on how children engaged with AI Puzzlers to critically reflect on genAI's capabilities. After being introduced to the system, children were divided into five groups and encouraged to collaborate, discuss strategies, and work together to solve puzzles in manual mode. This allowed them to familiarize themselves with the puzzle format and develop problem-solving strategies without AI assistance. Before introducing the AI mode, facilitators posed two key questions to prompt reflection: 1) "*Do you think genAI can solve these puzzles quickly or slowly? Why?*" and 2) "*Do you think genAI can solve these puzzles without any help from people?*" These questions aimed to capture children's initial expectations about genAI's capabilities before they engaged with it directly. Children then interacted with AI mode, where they could request AI assistance, observe AI-generated solutions, and receive AI-generated explanations for the puzzles. The session concluded with a 15-minute group discussion, where children shared their reflections on genAI's performance.

4.2.2 *Session 2.* Similar to Session 1, we began with a warm-up question: "*Tell us about a time you had to help others do something?*" This discussion encouraged children to reflect on how humans support one another, which facilitators then connected to the role of human guidance in helping genAI solve puzzles. Unlike Session 1, where children primarily observed genAI's independent performance, this session encouraged them to actively assist genAI by providing hints and experimenting with different strategies in Human-AI mode. Facilitators guided their exploration with key reflection questions such as, "*What hints do you think would be helpful?*" or, "*What do you think will happen if we change the number of examples?*" After 50 minutes of interacting with Human-AI mode, the different groups came together, and each team presented their experiences in front of the whole group, reflecting on successful strategies, challenges they encountered, and genAI's limitations.

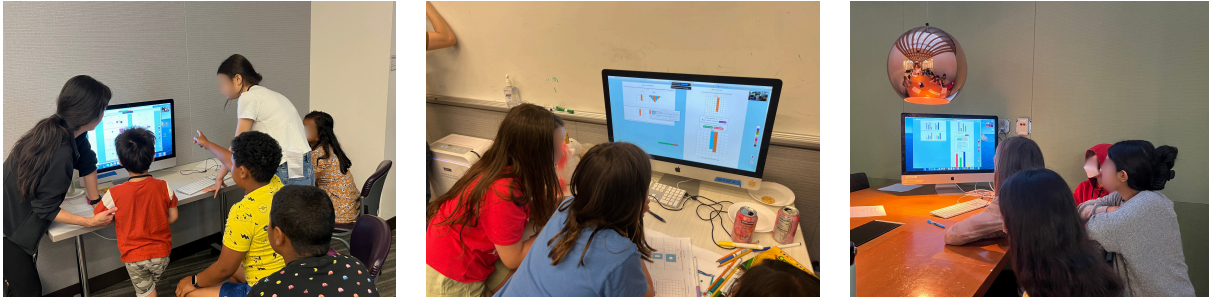

Figure 6: Children engaging with AI Puzzlers alongside adult facilitators.

## 4.3 Data Collection & Analysis

For both study sessions, our team used built-in webcams on desktop computers to record video and screen activity via Zoom, a video conferencing software. In total, we collected 927 minutes of video data. To capture additional insights, facilitators took field notes throughout the sessions, documenting key observations and notable moments. Additionally, physical artifacts, such as children's handwritten notes and collaborative sketches, were photographed for documentation.

The first, second, and fourth authors, then created analytical memos for all the videos [6, 66]. As part of this process, one author served as the primary reviewer, while another served as the secondary reviewer. The primary reviewer first watched the assigned recorded video and created a narrative summary of events at five-minute intervals. They documented children's interactions with AI Puzzlers, their reactions captured by the camera, and interactions and dialogues between participants, including direct quotes relevant to the study's research questions. After the primary reviewer completed their memos, the secondary reviewer independently reviewed the same videos and memos to verify the accuracy of the initial observations and to add supplementary insights. This dual-review process ensured the reliability of the data and captured a broader range of perspectives.

Following the creation and review of the analytic memos, we divided the memos into two equal-sized sets and began an inductive coding process [27]. The first two authors independently engaged in open coding of the first set of memos, suggesting potential codes such as "Interaction with the System" and "Making sense of AI." They then met over four meeting sessions to compare, reconcile, and refine the codes. During these discussions, they shared potential codes and their descriptions, collaboratively examined example quotes and counter-examples, compared code categories against one another, and refined the boundaries and definitions of each code. For example, the codes "Reading AI Explanations" and "Comparing Outputs with Correct Solutions" were consolidated into a single subcategory "Strategies." This iterative process led to the development of a codebook that included three main code categories: 1) Perception of AI, 2) Evaluation of AI Performance and 3) Interaction with the System. Once the codebook was finalized, the first author applied the final codes to the full dataset, and the second author conducted a second pass to ensure comprehensive analysis. We assessed interrater reliability through qualitative negotiations, where both authors met to discuss and resolve any coding disagreements [47]. We then organized the codes into overarching themes through two rounds of refinement and discussion. After finalizing the themes, the first author revisited the entire dataset to extract representative quotes for each theme, ensuring that the themes were well-supported by the data.

## 5 Findings

We present our findings from children's interactions with AI Puzzlers, focusing on how they engaged with the system, responded to genAI's successes and failures, and developed an understanding of its capabilities and limitations. To illustrate their learning process, we use representative vignettes, embedding our analysis within each example. While this study examines children's learning about generative AI, they generally referred to genAI as "AI" in their discussions. For consistency with their language, we use "AI" throughout this section while maintaining an analytical focus on genAI's reasoning processes.

### 5.1 Children's Interest and Exploration of AI Puzzlers

*5.1.1 Surprise, Excitement, and AI's Unexpected Errors.* At the start of Session 1, when we first introduced children to AI Puzzlers, they had high expectations that AI could solve the puzzles. This belief in AI stemmed from their own ease in solving the puzzles, their perceived belief in AI's ability to use visual references to recognize patterns, and their general trust in AI's broad knowledge. However, they quickly noticed that AI struggled to solve the puzzles and their reactions, verbal and physical, reflected their surprise.

For example, in Session 1, when Ivy, Ren, Emi, and Mara first interacted with AI Puzzlers, Emi eagerly volunteered to solve the second puzzle (see Figure 7). She identified the pattern as "*alternating between red and grey*" and correctly solved the puzzle. When the facilitator asked about the puzzle's difficulty, the group quickly agreed that it was simple. Ivy explained, "*It was pretty easy, judging by all the pictures. It was kind of obvious what you have to do since [the pattern is] that color, grey, that color, grey.*" Ren added, "*It can be easily identified what the pattern is going to be.*"

When the facilitator asked if AI could solve the puzzle, Emi, Mara, Ren, and Ivy all agreed that it could because "*It was easy.*" Ivy reasoned, "*There are tons of different references. Judging by all the references, since AI takes references from other pictures, the reference it gets is that color, grey, that color, grey.*" Here, Ivy's comment implies that she believed AI could use abstraction skills by generalizing from specific examples "*the references*" to form a rule (alternating between the color and grey) needed to solve the puzzle. Mara added, "*Because AI kind of knows everything,*" indicating that she assumed AI's vast access to information would enable it to solve the puzzle. Her response revealed a gap in her understanding of AI's limitations. However, when the AI returned an incorrect solution (see Figure 7), the group burst into laughter. They found the situation humorous because the AI failed so badly, despite their prior confidence that it would succeed, making the error an unexpected source of amusement. Additionally, the visual nature of the puzzles allowed them to quickly recognize the mistake, as Ivy commented, "*That is very very wrong.*"

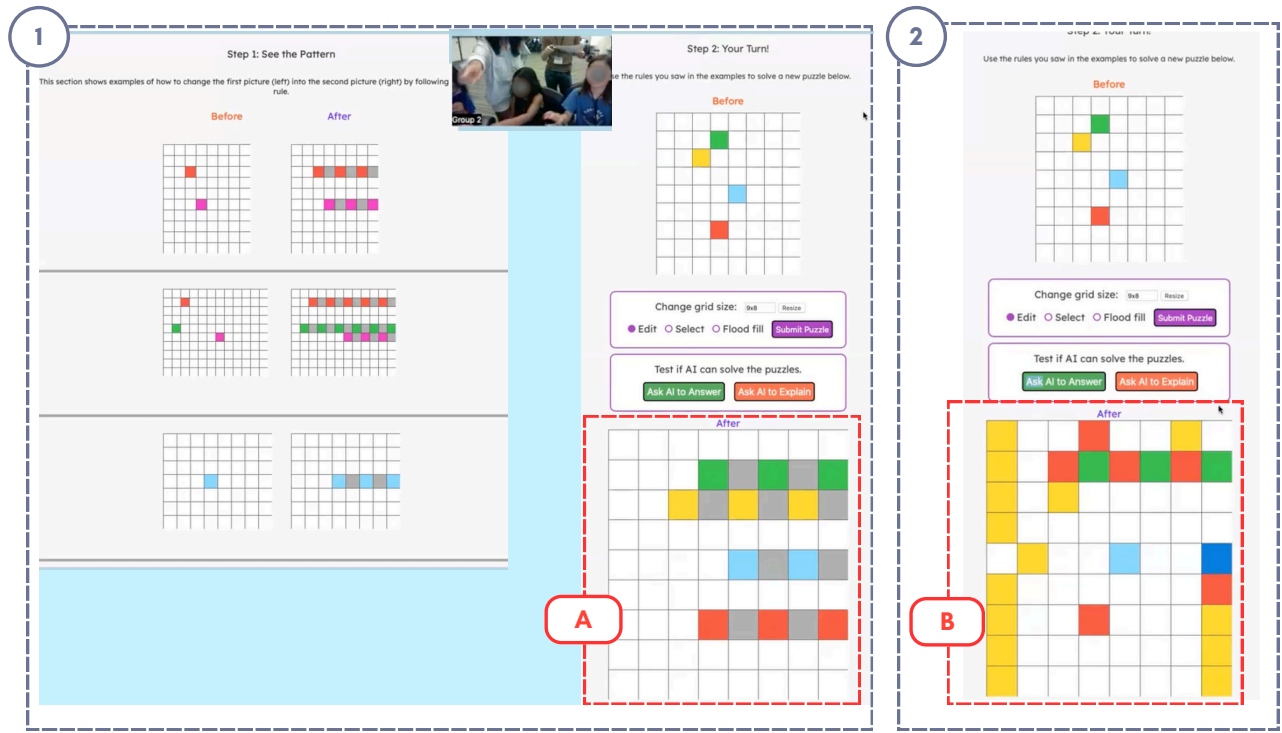

Figure 7: Panel (1) shows children collaboratively solving a puzzle, while panel (2) presents the AI's attempt at the same puzzle. (A) highlights the children's correct solution, whereas (B) shows the AI's incorrect attempt.

The sharp contrast between the AI's solutions and the children's correct answers continued to evoke surprise and amusement, even after they had repeatedly watched the AI fail. For instance, midway through Session 1, Juno and Hana had seen the AI struggle with nine previous puzzles, all of which they had solved correctly. Reflecting on the AI's past failures, Hana predicted, "*Maybe it can solve the puzzle faster but incorrectly*" as they prepared to ask the AI to solve the tenth puzzle (see Figure 8). Yet, when the AI returned another incorrect solution (see Figure 8), Juno said, "*Oh my gosh. What the heck is this?*" Hana, amused, blew a raspberry and added, "*It makes no sense whatsoever,*" to which Juno agreed, saying, "*It did something very weird.*"

Moreover, children's engagement with AI Puzzlers wasn't solely tied to the AI's mistakes; they were also drawn in by the puzzles themselves. Children viewed solving the puzzles as opportunities for problem-solving and personal accomplishment. Even as the puzzles became increasingly difficult, the children maintained their interest, often describing the more complex puzzles as "*fun.*" Peer collaboration consistently reinforced this enthusiasm, as children confirmed and validated each other's solutions before turning to the AI for comparison. For example, in Session 1, Juno and Hana, eagerly tackled the puzzles from the start. Looking at their next puzzle (see Figure 9) Juno remarked, "*This is pretty easy,*" while Hana added excitedly, "*I want to do this because it looks so fun.*"

Together, they analyzed the puzzle, with Hana reasoning through the height and color relationships, stating, "*It's the tallest one and the smallest one, correct? Blue would be the tallest.*" Juno confirmed, "*This looks right,*" and the pair successfully solved the puzzle while the AI failed (see Figure 9). As their excitement grew, Hana exclaimed, "*We can go forever.*" The facilitator reminded them, "*There's only up to level 4,*" to which Juno confidently replied, "*We can go farther than the AI,*" and Hana affirmed, "*Yes, we can.*" This exchange highlights the children's belief in their abilities, expressing confidence in surpassing the AI. Reflecting on their success, Hana laughed and said, "*AI would have gone out at the first question,*" underscoring their shared sense of accomplishment and enjoyment in outsmarting the AI.

Overall, two key dynamics drove children's sustained engagement with AI Puzzlers in Session 1: 1) their surprise at the AI's mistakes, especially for puzzles they considered "easy," and 2) the satisfaction of solving challenging puzzles and comparing their correct solutions to the AI's. Despite the AI's repeated failures, children eagerly engaged with each puzzle, enjoying the process of outsmarting the system, testing their problem-solving skills, and comparing solutions. This blend of competition, collaboration, and humor kept them motivated, creating an engaging experience.

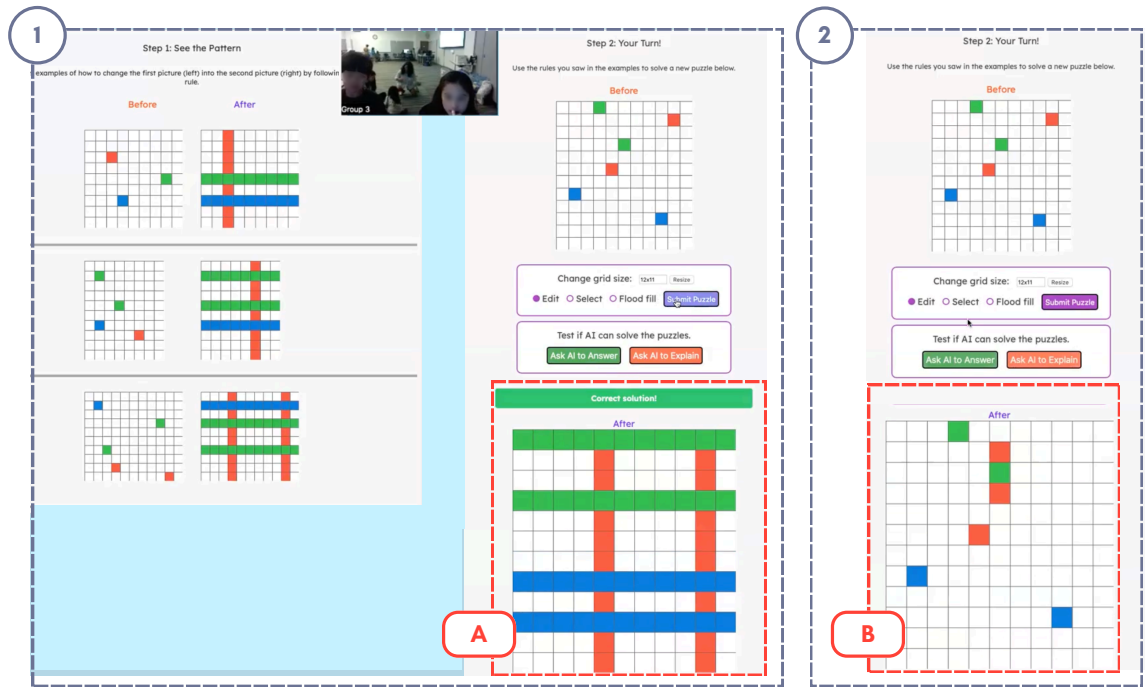

Figure 8: Panel (1) shows children collaboratively solving a puzzle, while panel (2) presents the AI's attempt at the same puzzle. (A) highlights the children's correct solution, whereas (B) shows the AI's incorrect attempt.

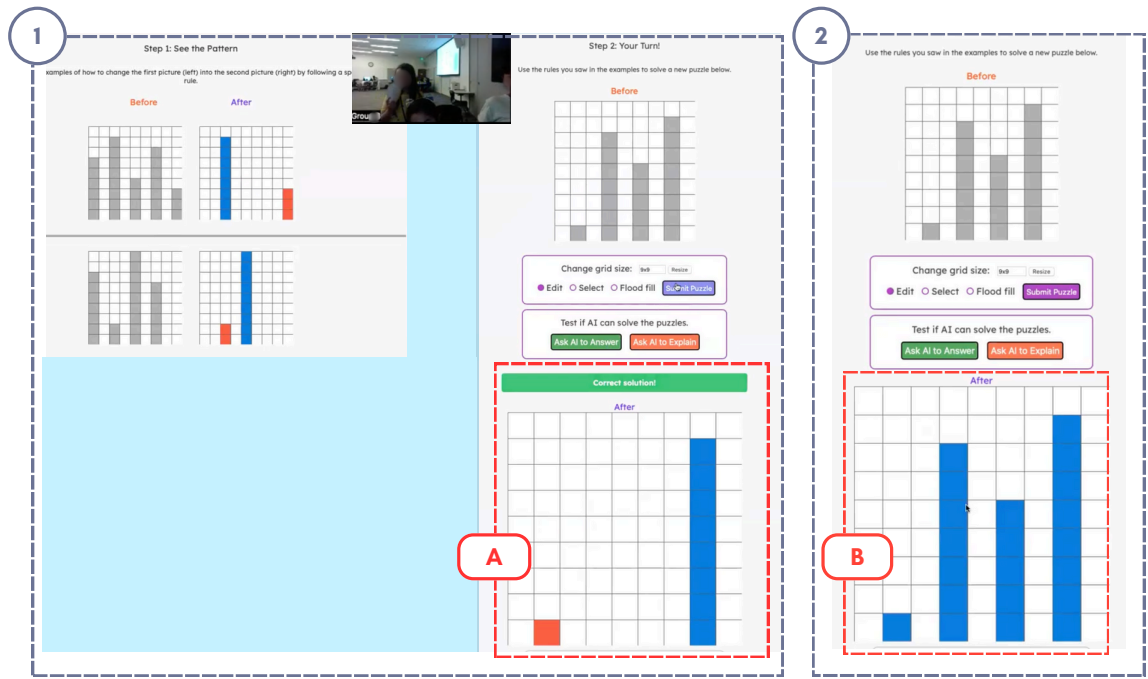

Figure 9: Panel (1) shows children collaboratively solving a puzzle, while panel (2) presents the AI's attempt at the same puzzle. (A) highlights the children's correct solution, whereas (B) shows the AI's incorrect attempt.

*5.1.2 Children's Iterative Debugging of AI Errors.* As children interacted with AI Puzzlers, they quickly spotted incorrect AI solutions due to the visual nature of the puzzles, and more importantly, understood how the AI was failing. Similar to debugging in programming, in Session 2, children actively guided the AI by identifying its errors, providing corrective hints, testing their hints, and refining their instructions based on the AI's outputs. While the visual modality of the puzzles made it easier for the children to iteratively test and improve their instructions, they also encountered AI's limitations, realizing that it often misunderstood or failed to fully follow their instructions.

For example, in Session 2, Ivy, Juno, Rina, and Elise were trying to help the AI solve the second puzzle (see Figure 10). The children's debugging process began when Ivy suggested the first hint, "*Make a pattern of gray and a different color,*" while Juno proposed a more complex version: "*Make a pattern of gray, the color gray's after, and then gray, and the color gray after.*" After some discussion on avoiding confusion for the AI, the group settled on a simpler hint, "*Make a pattern of the colors and gray.*" However, when the AI produced an incorrect grid of colors and gray, Rina immediately recognized the mistake, commenting, "*It still can't solve it though.*" The children examined the AI's output and refined their instruction to, "*Make a pattern of the colors and gray alternating,*" which led Rina to observe that the AI's output was "*kind of alternating now.*" At this stage, the children recognized the AI's partial success but also noticed that it didn't fully capture the pattern they had envisioned. Ivy pointed out another issue with the background color, saying, "*Now we need to add a background of white.*"

The children's increasing specificity in their hints mirrored the process of narrowing down AI's errors by fine-tuning instructions. Ivy specified the background color in her instruction, typing "*make a pattern of the colors and gray alternating and a background of white,*" but when the AI's output still didn't reflect the correct background, Ivy remarked, "*That's not a background of white.*" The children had to systematically address different parts of the puzzle—first the pattern, then the background—while the AI continued to misinterpret or overlook aspects of the instructions. On their second attempt with the same hint, the AI came very close to solving the puzzle, prompting a celebratory "*Yayyy!*" from the group. However, Ivy quickly commented "*There we go, but it forgot the yellow,*" while Rina noted that "*AI's answer doesn't always include all the colors shown in the original puzzle question.*" This led the group to modify their hint again, including a list of all the colors: "*Make a pattern of the colors and gray alternating and a background of white, red, light blue, green, yellow.*" By listing all the colors, the children demonstrated an understanding that more explicit, step-by-step guidance might help the AI avoid leaving out key elements. However, despite their efforts to refine and clarify their instructions, the AI continued to misinterpret them, leading Ivy to comment, "*I am so done with you, AI.*" Ivy's expression of frustration reveals her realization that, despite the group's efforts to provide increasingly explicit and refined instructions, the AI was still unable to fully grasp their intent.

Despite recurring misunderstandings, children not only refined their hints but also, in several instances, brainstormed alternative strategies to overcome the AI's limitations. For example, in Session 2, Selah, Emi, Ren, and Mara attempted to help the AI solve the third puzzle (see Figure 11). The children's debugging process first involved discussing what specific directions they should give the AI. Emi typed the hint, "*surround the colors with,*" before pausing. Ren then suggested, "*I think we shouldn't say yellow, purple, and blue...I think we should say something simple.*" Taking this into account, Emi simplified the instruction to, "*surround the colors with more colors.*" However, the AI failed to generate a correct output, prompting Ren to suggest a new approach, "*What if we ask it to make donut shapes around the colors?*" Ren's suggestion reflects an attempt to simplify the instruction by using a familiar visual metaphor—a donut shape—to describe how the colors should be surrounded by rings or circles. The children likely believed that this metaphor would make sense to the AI since it's based on a familiar object from their everyday experiences.

Selah then typed the revised hint, "*make a donut shape around the color*" (see Figure 11). However, even with this more specific instruction the AI still couldn't produce the correct result. This illustrates a key limitation in how the AI fails to interpret visual metaphors that are culturally relevant or based on children's experiences. While the concept of a "*donut shape*" is clear to the children, the AI struggles to grasp such human-centered metaphors. In discussing design ideas on how to resolve AI's issues, Emi suggested if they can test whether the AI could understand human instructions paired with visuals. She began drawing the correct solution on the grid to show what it would look like. To the facilitator's question, "*Do you think if [we] gave it human instructions with pictures it would understand?*" Mara and Selah responded, "*Yeah,*" optimistic that the visuals might help bridge the gap. While the AI Puzzlers system didn't have the functionality to process both visual and textual hints, we see this as an example of how the children not only iteratively refined their approach to guide the AI but also came up with a new idea to interact with the AI as they confronted its limitations.

Overall, children's interaction with AI Puzzlers in Session 2 demonstrated their ability to actively engage in a process of problem-solving and debugging. Through iterative refinement of their hints, children adapted their strategies, and even explored creative approaches like using visual metaphors or pairing instructions with potential visual aids. This highlights both their resilience and their capacity for critical thinking in navigating AI's limitations. However, their engagement was not without challenges. The AI's persistent misunderstandings, even after detailed and increasingly refined instructions, sometimes led to frustration. This underscores the difficulty of working with an AI system that consistently failed to meet their expectations, showing how extended interactions with imperfect AI systems can affect children's engagement.

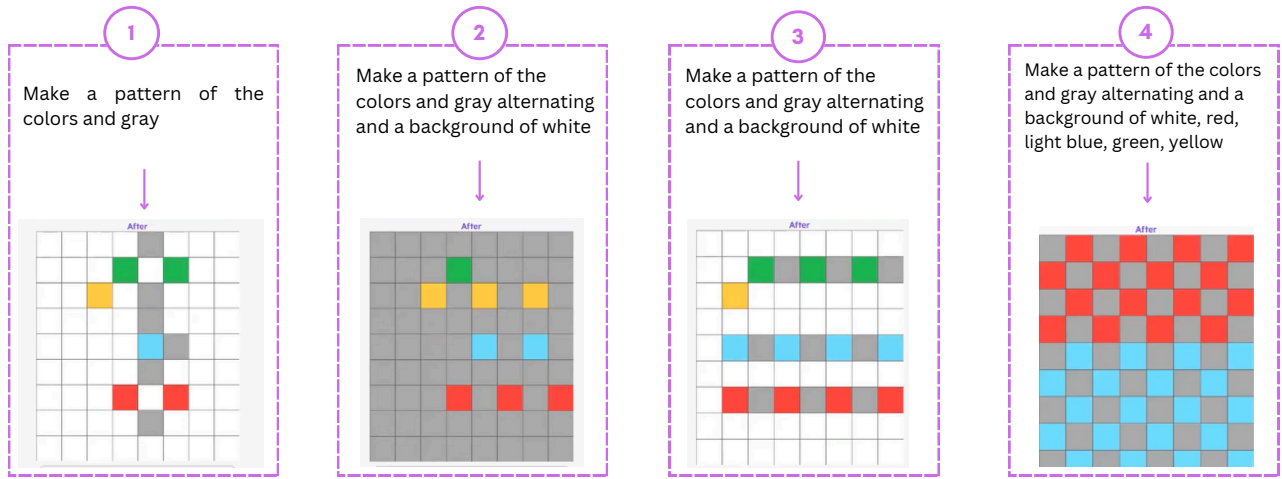

Figure 10: Children iteratively refined their instructions to guide the AI towards solving the puzzle. The sequence showcases the increasing specificity in their hints and AI's corresponding outputs.

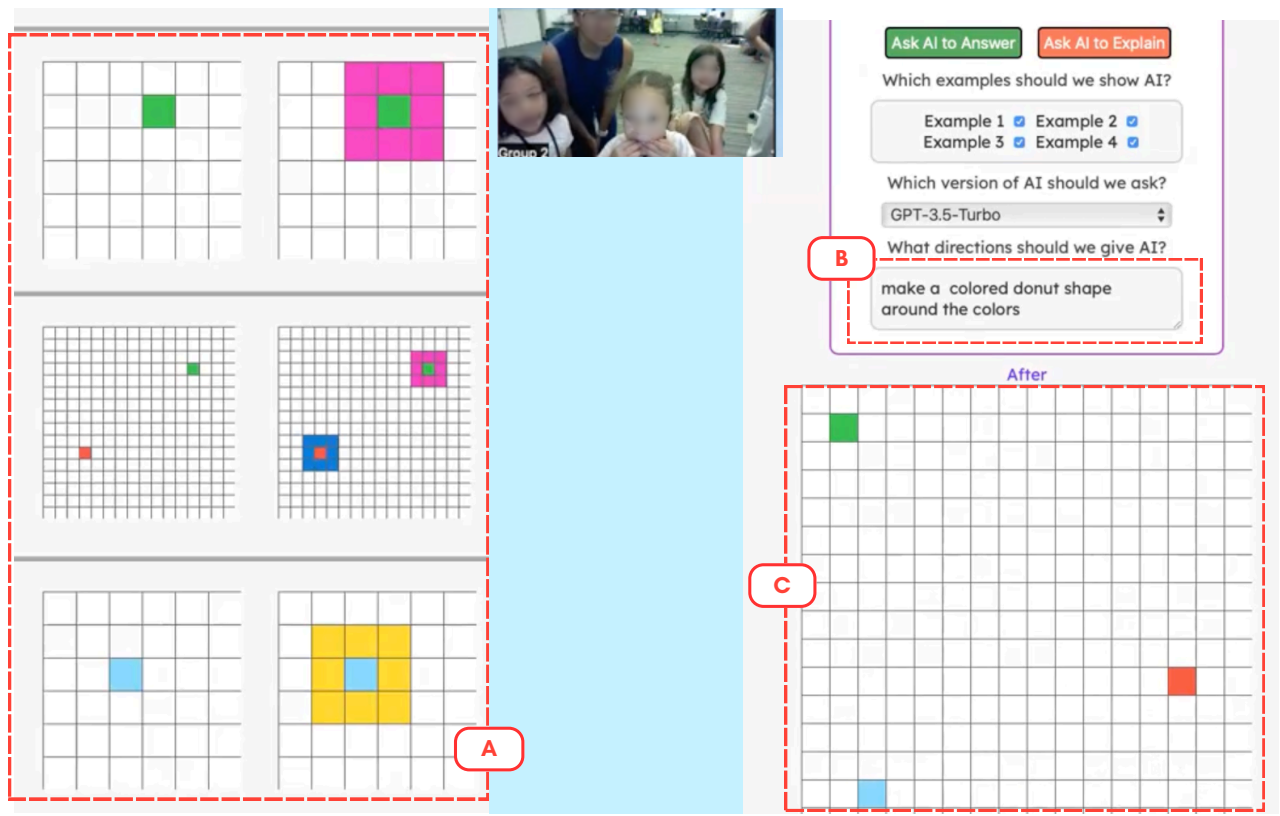

Figure 11: Children's Interaction with Assist Mode in AI Puzzlers. (A) Displays examples of example patterns to infer the transformation rule. (B) Shows the instruction children provided to the AI: "make a colored donut shape around the colors." (C) Illustrates the AI's generated output in response to their instruction.

## 5.2 Understanding AI's Limitations Through Observing Inconsistencies

*5.2.1 Children Identify Inconsistencies in AI's Reasoning.* While interacting with AI Puzzlers during Session 1, children reflected on AI-generated outputs by cross-examining the AI's visual solutions with its explanations. There were many examples of children identifying discrepancies between the AI's reasoning and its actual puzzle outcome. For example, in Session 1, after Hana and Juno correctly solved a puzzle, they asked the AI for its solution. Upon reviewing the AI's answer, they quickly identified it as incorrect and requested an explanation from the AI (see Figure 12).

After reading the AI's explanation, Hana remarked, "*This*"—pointing at the explanation—"*has nothing to do with this*,"—pointing at the AI's pattern. She then added, "*It's like someone who is not listening,*" suggesting that she perceived the AI's explanation as disconnected from its visual output, much like a human who isn't paying attention. Juno built on this observation, saying, "*The AI said it was following a pattern, but what pattern is this? This is not a pattern. It is just red, red, red,*" as he pointed at the AI's solution. This example demonstrates that the children were not simply accepting the AI's reasoning—they were actively engaging with its logic, looking for coherence between explanations and visual results. Furthermore, having correctly recognized the correct pattern earlier, they saw that the AI's solution was unrelated to the pattern they knew was correct, leading Hana and Juno to further question the AI's reasoning.

This type of critical engagement with AI's reasoning recurred throughout Session 1. Children frequently encountered inconsistencies not only between the AI's explanations and its visual outputs but also contradictions within the AI's explanations themselves. For instance, when Noa, Finn, Leila, and Zayn asked the AI to explain its solution (see Figure 13), the AI explained, "*I changed all the middle white cells to teal while keeping the corners white. This makes the grid look like a frame of teal around a white center.*" Finn considered the explanation and said, "*The explanation is right; it just did it wrong,*" implying that while the AI's reasoning was correct, the error was in its visual execution. However, Noa identified a contradiction between the AI's first and second sentences. "*It says the corners are white, but the corners should be teal,*" she pointed out, noting that if the corners were indeed white, as the AI claimed, the grid would not form the intended frame of teal. This highlights how children, while able to easily detect errors in visual outputs, had to engage in more critical reasoning to spot contradictions in AI-generated text-particularly when the text appeared superficially reasonable.

Similarly, in another group, the children recognized a comparable issue of superficial correctness in AI's reasoning where the AI's explanation seemed logical on the surface but did not provide enough detail for the children to understand its approach (see Figure 14). For instance, when Ivy, Mara, Ren, and Emi asked the AI to explain its reasoning for the second puzzle, the AI's explanation stated that it identified a pattern in how the colors changed in each row of the input grid. It claimed to have applied the same pattern to predict the outputs in the new puzzle. However, AI failed to specify the exact nature of the pattern or how the pattern influenced the color changes in the output, leaving critical parts of its reasoning vague.

This prompted Ren to remark, "*Well it is explaining, but I didn't understand it,*" highlighting the gap between the AI's surface-level explanation and the children's need for a more detailed and comprehensible breakdown of its reasoning. Ivy critiqued this vagueness saying, "*AI is very scientific, given its scientific explanation, but sometimes it's better not to go super, duper scientific and just go by your three references.*" By "*scientific,*" Ivy meant that the explanation was too technical or abstract lacking practical clarity. Her reference to AI should follow the "*three references*" reflects her preference for the AI to apply a more straightforward approach to solving the puzzle—one that directly considers the three examples demonstrating the transformation rule, favoring a more direct and understandable reasoning.

Overall, children's critical evaluation of AI-generated outputs demonstrated their active role in scrutinizing the reasoning behind the AI's decisions. Rather than passively accepting the AI's responses, they engaged in a process of deeper analysis, identifying inconsistencies and contradictions by cross-examining the AI's visual solutions with its textual explanations. This critical reasoning went beyond merely checking if the answer was correct; it involved questioning the logic, coherence, and transparency of the AI's problem-solving process. Additionally, they sought to understand how the AI arrived at its conclusions, recognizing when explanations lacked detail or coherence, and pushing back when reasoning was unclear. These findings highlight the importance of designing AI systems that allow children to not just receive answers but also critically engage with the logic behind them.

*5.2.2 AI's "Scientific Brain" vs. Human Problem Solving.* Throughout Session 1, children recognized that AI approached problem solving differently from humans. While children found the puzzles "easy," they realized the puzzles were "super hard" for AI as it struggled to solve them, displaying inconsistencies in its reasoning. In several instances, children reflected on the traits of human problem-solving, such as reasoning, abstraction, and creativity—qualities they felt the AI lacked.

For example, in Session 1, after Noa, Leila, Finn, and Zayn correctly solved a puzzle, they asked AI to do the same (see Figure 15). The AI provided an incorrect solution, which prompted the group to ask AI again. On AI's second attempt, Finn observed, "*It figured out the pattern but didn't get the color positions right.*" Reflecting on this improvement, Finn hypothesized, "*AI is learning.*" This led the group to predict that the AI would perform better on the next attempt. However, on the third try, AI's solution was even worse (see Figure 15). This led Zayn to observe, "*AI doesn't have the same mind as us.*" This statement marked a turning point in the group's understanding of AI, as they began to distinguish between human cognition that is capable of learning and AI's processes. Zayn's observation prompted the facilitator to ask, "*Well, what's the difference in our minds?*" Leila noted, "*This is the internet's mind.*" Noa elaborated, "*It's trying to solve it based only on the internet, but the human brain is creative. AI only has the info it's given, but humans have other experiences.*" Their reflections highlight how children grasped the limitations of AI, understanding that AI's responses are constrained by the data it has access to such as the internet. By contrast, when Noa says, "*human brain is creative*" and "*humans have other experiences,*" she underscores how humans can draw on a wide range of knowledge, including emotional and situational experiences, to approach challenges more flexibly, whereas AI is limited to the information it has been given or can access.

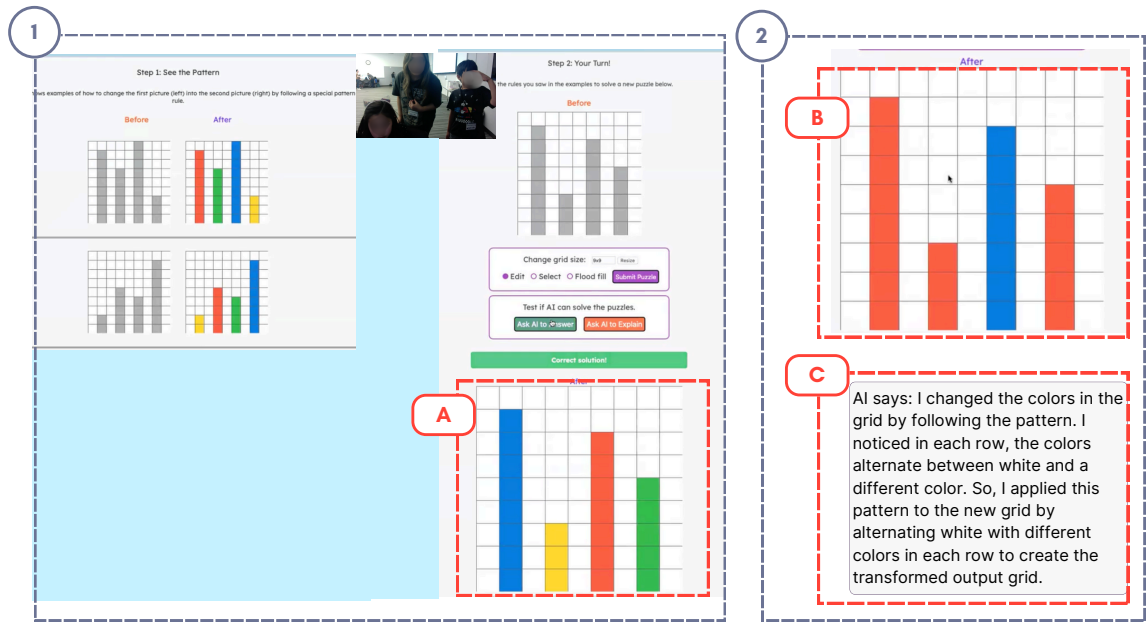

Figure 12: Panel (1) shows children collaboratively solving a puzzle, while panel (2) presents the AI's attempt at the same puzzle. (A) highlights the children's correct solution, whereas (B) shows the AI's incorrect attempt. (C) provides the AI's explanation of how it solved the puzzle.

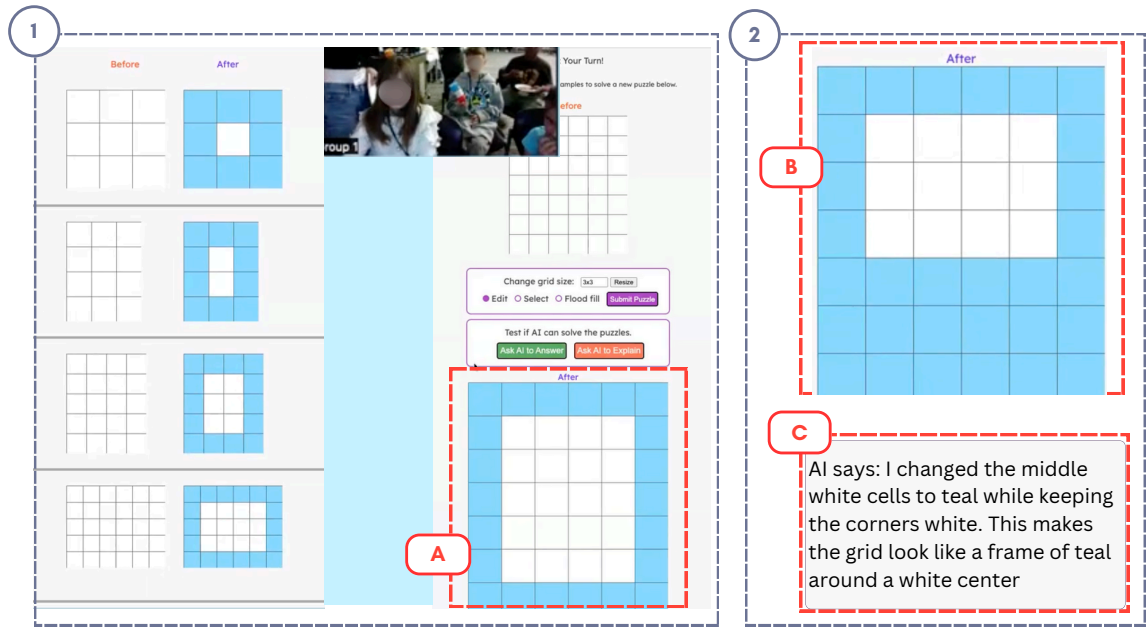

Figure 13: Panel (1) shows children collaboratively solving a puzzle, while panel (2) presents the AI's attempt at the same puzzle. (A) highlights the children's correct solution, whereas (B) shows the AI's incorrect attempt. (C) provides the AI's explanation of how it solved the puzzle.

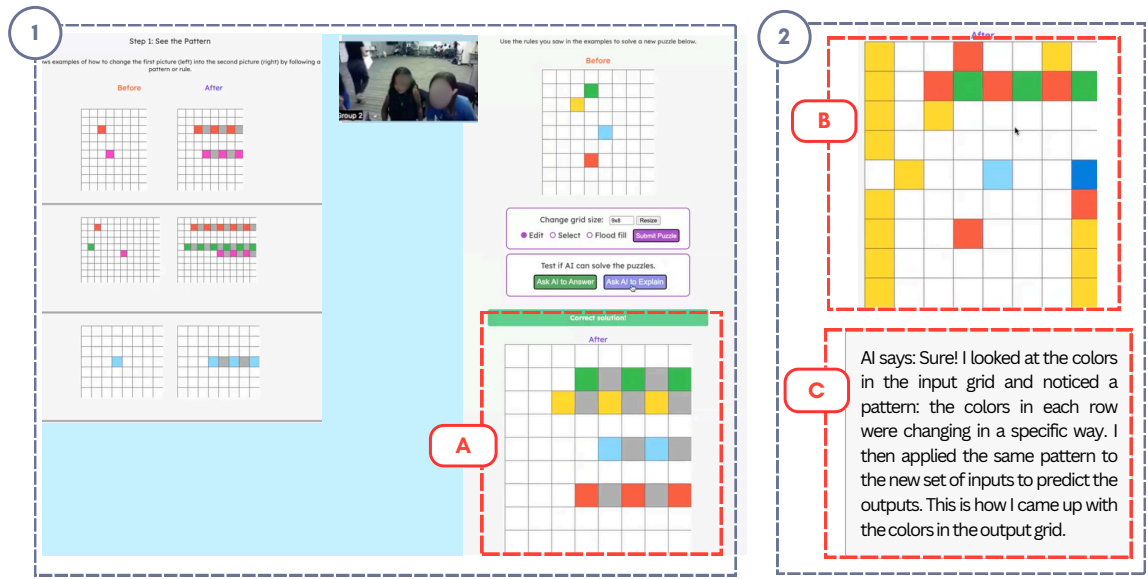

Figure 14: Panel (1) shows children collaboratively solving a puzzle, while panel (2) presents the AI's attempt at the same puzzle. (A) highlights the children's correct solution, whereas (B) shows the AI's incorrect attempt. (C) provides the AI's explanation of how it solved the puzzle.

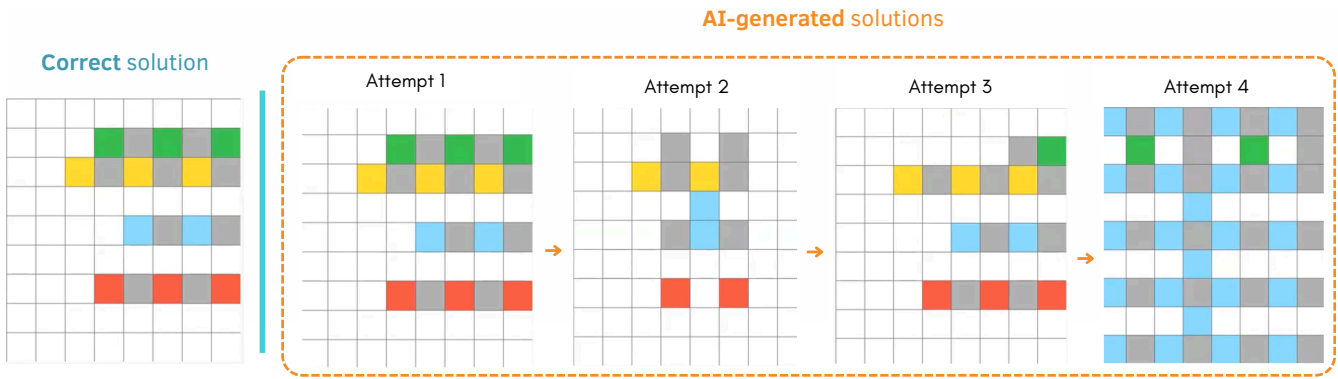

Figure 15: The correct solution (left) is compared with AI-generated attempts (right).

Another example of children differentiating between human and AI problem solving occurred in Session 1, when Ivy, Ren, Emi, and Mara observed the AI unsuccessfully attempting to solve a puzzle (see Figure 14). Pointing to AI's solution, Ivy asked, "*What is that blue? Look at the references and think like a human being!*" Here, Ivy was commenting on the AI's inability to abstract information from the references, as a human might, to solve the puzzle. Later, when the facilitator asked what they thought was happening to AI, Ren explained, "*It is taking the references and is able to copy and paste colors but not use the context of the patterns.*" Ren's statement highlights their understanding that, while the AI could process the data it was given (the references and colors), it was failing to apply abstraction and reasoning necessary to complete the task. Thus, the AI could mimic certain elements, such as color, but it struggled with the more complex, abstract reasoning needed to solve the puzzle accurately. Ivy added to this by saying AI's "*scientist's brain*" has "*ones and zeroes[that] help it understand the color, but ones and zeroes aren't the smartest.*" Her observation, along with Ren's, reflects the children's growing awareness that while AI can handle tasks like identifying colors, it lacks the creative and abstract reasoning necessary to grasp the puzzle's full context.

Similarly, children also observed the lack of reasoning in the AI's approach, describing its responses as "rapid random guessing." For example, in Session 1, when the facilitator asked Juno and Hana to reflect on how the AI solves the puzzle, Juno repeatedly clicked the "Ask AI to Answer" button, pointing out, "*Look, it changes every time.*" (see Figure 16).

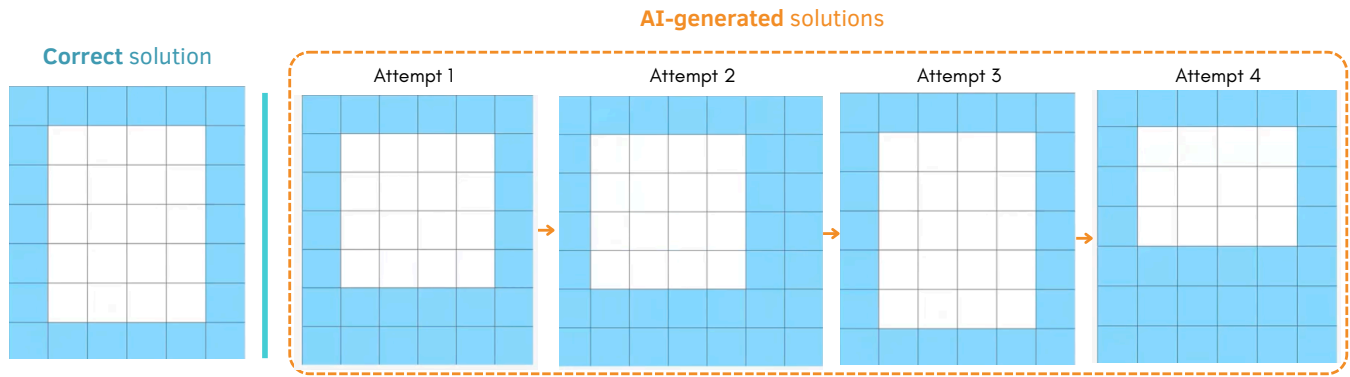

Figure 16: The correct solution (left) is compared with AI-generated attempts (right).

AI's lack of reasoning became evident to the children when Juno noted that the AI's answers changed with each attempt, implying no continuity in its problem-solving strategy. Instead of logically building on previous attempts, the AI appeared to reset its process, attempting a new guess each time without learning or adapting from prior mistakes. This behavior led Hana to remark "*AI just keeps guessing,*" and Juno added, "*AI is stupid*" and only "*gets lucky.*" When the facilitator asked, "*What can the AI do to be smarter?*" Juno responded, "*Not guess,*" indicating the children's growing awareness that the AI was not reasoning but merely stumbling upon the correct answer by chance.

Overall, children's reflections across Session 1 and Session 2 demonstrated their growing awareness of the fundamental differences between human and AI problem-solving. They recognized that AI operates within rigid parameters, relying on data it has access to, often defaulting to trial-and-error rather than employing the kind of reasoning and creativity characteristic of humans. Their experiences reinforced the notion that AI, while capable of processing and replicating information, lacks the ability to apply abstraction and draw on diverse experiences to solve reasoning problems.

## 6 Discussion

In this section, we discuss how our findings connect to prior literature on AI literacy and explore their implications for child-AI interaction. Specifically, we consider how AI Puzzlers positioned children in active, inquiry-driven roles that encouraged experimentation and critical reflection on generative AI's reasoning processes. Our findings offer recommendations for designers of AI systems for children, as well as researchers and educators who support children's engagement with AI technologies.

### 6.1 Positioning Children as Active Inquirers in GenAI Interaction

AI Puzzlers positioned children as active inquirers, encouraging them to identify, analyze, and debug errors in genAI's outputs. By embedding this process in puzzle-based gameplay, the system sustained engagement while fostering reflection on AI's limitations and capabilities. Below, we elaborate on each of these aspects.

*Encouraging Critical Evaluation of GenAI's Outputs.* From the outset, children expressed confidence in AI's ability to solve ARC Puzzles, reflecting a general belief in AI's capabilities. This belief aligns with broader patterns observed in prior research, where children's early interactions with AI technologies often reflect an overestimation of AI's intelligence [2, 49, 62, 77]. However, as they engaged with AI Puzzlers, children quickly realized that the puzzles they considered "easy" were challenging for the AI. The puzzles' visual nature ensured that there was no obscurity in the way genAI presented information, while accompanying textual explanations provided a step-by-step breakdown of the AI's reasoning [44, 46]. This led to moments of visible disbelief, such as when children reacted to genAI's incorrect solutions with exclamations like "WHAT!!"

*Encouraging Iterative Debugging & Reflection.* The Assist Mode in AI Puzzlers–where children provided hints to guide the AI–helped them develop an emergent schema of genAI's capabilities and limitations. As they iteratively refined their hints, shifting from vague commands like "make a pattern" to more precise instructions like "make a pattern with alternate colors," they demonstrated a growing understanding of how genAI processes information. The persistence children showed in in fine-tuning AI's outputs suggest that they viewed AI as a system requiring guidance rather than an infallible problem solver. Notably, we found that children's explanations of AI's reasoning became more sophisticated when using Assist Mode compared to AI Mode, where they only observed the AI making mistakes. These findings align with prior research, emphasizing the importance of providing children with opportunities to actively experiment with AI models and debug their assumptions about AI. [22, 35, 36, 50].

*Fostering Sustained Engagement Through Puzzles.* The game-like nature of the puzzles revealed children's inclination to engage in competitive problem-solving, often positioning themselves against the AI in an effort to "beat" its performance [38, 55, 97]. This dynamic, which aligns with game-based learning and the motivational role of competition in games [12, 17, 82, 98], not only sustained engagement but also reinforced children's recognition of their own problem-solving strengths. Prior research suggests that children often perceive AI as lacking in creativity and flexible thinking [43], and similar patterns emerged in AI Puzzlers. As children saw that the AI struggled with the ARC puzzles, they began comparing how humans and AI solve problems. This led them to recognize AI's reasoning limitations and also made them more confident in their own abilities. This supports Long and Magerko's [41] argument that fostering an understanding of both AI's limitations and human strengths can empower users to leverage their own cognitive abilities in domains where AI falls short. In the next section, we present design implications for supporting AI literacy, grounded in these insights.

## 6.2 Implications for AI literacy and Generative AI Systems for Children

*6.2.1 Designing for Interpretability Without Cognitive Overload.* Building on prior research on AI interpretability [33, 37, 63], we emphasize the importance of making AI decision-making understandable for fostering AI literacy. While generative AI models can provide justifications for their outputs, overly lengthy or text-heavy explanations may overwhelm young users and discourage engagement [69]. Future genAI systems could support interpretability by generating visual reasoning traces. For example, genAI systems could create real-time visual representations, such as decision trees, flowcharts, or animated characters that walk through the AI's reasoning process. This would allow children to trace AI's logic step by step, encouraging them to question or revise AI-generated reasoning. Additionally, in AI Puzzlers, we found that children benefited from side-by-side visual and textual comparisons, which helped them verify correctness before engaging with the AI's explanation. To distribute cognitive load across visual and verbal channels [44, 46], AI-generated explanations could highlight or animate key visual elements, with corresponding narration in real time to reinforce its reasoning.

*6.2.2 Using Validity Markers to Guide Children's Attention.* Inspired by prior work on human-AI collaboration [37, 81], we argue that AI systems designed for children should not only present information but also support them in evaluating its reliability and accuracy. Research on uncertainty visualization [7, 60] demonstrates that well-designed visual markers can help users, including children, focus on areas requiring deeper scrutiny. Building on this idea, genAI tools for children can incorporate explicit validity markers, such as color-coded confidence levels (e.g., green for high confidence, orange for uncertainty), uncertainty flags (e.g., warning icons next to questionable AI-generated answers) or playful prompts (e.g., "Try testing this answer!"), to prompt critical attention and active evaluation. However, prior studies suggest that such markers must be carefully designed to prevent over-reliance on AI confidence indicators, as children may assume high-confidence outputs are always correct. Thus, validity markers could be paired with explicit scaffolds that support evaluation strategies, such as comparing alternative solutions, identifying contradictions, testing claims with counterexamples, and justifying answers with evidence.

*6.2.3 Designing for Reflection Through AI Experimentation.* Prior research suggests that children develop a deeper understanding of AI concepts when they can tinker with system parameters and observe how different inputs influence outcomes [10, 19, 26, 65]. In AI Puzzlers, children engaged in tinkering and experimentation by adjusting the number of examples provided to AI, modifying hints, and selecting different AI versions to see how these changes influenced AI-generated solutions. This process was not solely about obtaining correct answers but also about interrogating AI's reasoning and making sense of its decision-making patterns. Similarly, AI systems can be designed such that instead of simply watching an AI generate an explanation, a child might be able to pause at key points, tap on specific parts to get additional details, or even manipulate variables to see how AI reasoning changes. However, we also observed that children sometimes struggled to interpret AI's reasoning, particularly when multiple factors influenced an output. This suggests that AI literacy tools should not only support experimentation but also scaffold reflection on how AI arrives at its responses. For example, AI-enabled platforms could allow children to modify AI parameters while also providing structured prompts that encourage them to articulate hypotheses, compare outcomes, and reflect on differences. By embedding both interactive tinkering and guided reflection, these systems can support both "reflect in action" (thinking while doing something) and "reflect on action" (thinking after you have done it) [67].

## 7 Limitations & Future Work

While our study's approach was designed to maximize depth and rigor, certain methodological choices naturally shaped the scope of our findings and suggest directions for future work. First, our co-design sessions engaged 21 children from a single geographic region, all of whom had prior experience with participatory design. This experience likely shaped their ability to confidently share opinions with adults, engage in constructive disagreements, and articulate their reasoning in real time. While this facilitated rich discussions, it also means that our findings should be understood as formative theoretical generalizations rather than statistical generalization [91]. Future work could examine how children in different settings, such as schools and libraries, and across diverse cultural and linguistic contexts engage with AI Puzzlers. Expanding to these contexts would help illuminate how different social, educational, and cultural dynamics shape children's evaluation of genAI outputs and support a broader understanding of AI literacy development globally.

Second, although our findings show that children can detect and reflect on genAI errors within the structured environment of AI Puzzlers, we did not assess whether this learning transfers to other, more open-ended genAI interactions. Future work will involve follow-up studies to examine this transferability, investigating whether engaging with AI Puzzlers supports children in interpreting, questioning, or critiquing genAI systems they encounter in real-life contexts. We also plan to expand AI Puzzlers to include other modalities, such as voice-based interactions to more fully represent the range of AI systems children encounter in their daily lives.

Third, we selected ARC Puzzles because they are both engaging for children and commonly used as a benchmark for evaluating AI reasoning [14]. At the same time, we acknowledge that their reliance on color-based differentiation may pose accessibility challenges. Future work could build on our approach by exploring alternative puzzle formats or game-based approaches to broaden accessibility. Lastly, at the time of our study, we used the most advanced available version of ChatGPT (GPT-4o). Children also had opportunities to engage with four other versions of ChatGPT for understating variations in AI performance. As AI systems continue to evolve, their problem-solving capabilities will inevitably shift, and more recent models like OpenAI o3 (currently undergoing safety testing prior to release [57]) exhibit improved performance on the ARC puzzles [34]. Future work could extend AI Puzzlers by incorporating both newer and older genAI models to surface these shifts and examine how children recognize and interpret genAI's changing capabilities. This could also inform the design of educational tools that scaffold children's reflections on AI's evolving efficacy, supporting their ability to critically engage with AI technologies over time.

## 8 Conclusion

In this study, we presented AI Puzzlers, an interactive system designed to help children critically engage and analyze generative AI's outputs. Through participatory design sessions with 21 children (ages 6–11), we examined how they detected inconsistencies in genAI outputs, debugged AI-generated errors, and refined their strategies for guiding AI. Our findings underscore the need for genAI systems to present information in ways that support visual and textual comparison for reducing cognitive overload, foster active inquiry, and scaffold multiple ways of understanding AI-generated content. We hope that our work will inform the design of AI literacy tools that empower children to critically evaluate AI-generated content and develop a deeper understanding of AI's strengths and limitations.

## Acknowledgments

This material is based upon work supported under the AI Research Institutes program by the National Science Foundation and the Institute of Education Sciences, U.S. Department of Education, through Award #DRL-2229873 - AI Institute for Transforming Education for Children with Speech and Language Processing Challenges (or National AI Institute for Exceptional Education). Any opinions, findings, and conclusions or recommendations expressed in this material are those of the author(s) and do not necessarily reflect the views of the National Science Foundation, the Institute of Education Sciences, or the U.S. Department of Education. This work was also partially funded by the Jacob's Foundation CERES Network.

## 9 Selection & Participation of Children

Children who participated in our study were engaged in an intergenerational co-design group at our university. Parental consent and participant assent were obtained for every participant. Assent forms were written using an age-appropriate language. Consent and assent forms were approved by the Institutional Review Board (IRB) that reviews and oversees human subjects research in our institution. Parents and children were informed about the study's purpose, potential risks, and confidentiality measures. They were also assured that participation was voluntary, and children could withdraw at any time. During study activities, researchers acted as facilitators, ensuring that children did not feel pressured to participate. All adult facilitators completed institutional ethics and safety training for working with children. To protect participants' privacy, children's data was anonymized and securely stored.

## References

[1] Luca Ambrosio, Jordy Schol, Vincenzo Amedeo La Pietra, Fabrizio Russo, Gianluca Vadalà, and Daisuke Sakai. 2024. Threats and opportunities of using ChatGPT in scientific writing—The risk of getting spineless. *JOR Spine* 7, 1 (2024), e1296–n/a.

[2] Valentina Andries and Judy Robertson. 2023. Alexa doesn't have that many feelings: Children's understanding of AI through interactions with smart speakers in their homes. *Computers and Education: Artificial Intelligence* 5 (2023), 100176. doi:10.1016/j.caeai.2023.100176

[3] Zied Bahroun, Chiraz Anane, Vian Ahmed, and Andrew Zacca. 2023. Transforming education: A comprehensive review of generative artificial intelligence in educational settings through bibliometric and content analysis. *Sustainability* 15, 17 (2023), 12983.

[4] Emily M Bender, Timnit Gebru, Angelina McMillan-Major, and Shmargaret Shmitchell. 2021. On the dangers of stochastic parrots: Can language models be too big?. In *Proceedings of the 2021 ACM conference on fairness, accountability, and transparency*. 610–623.

[5] Gernot Beutel, Eline Geerits, and Jan T Kielstein. 2023. Artificial hallucination: GPT on LSD? *Critical Care* 27, 1 (2023), 148.

[6] Melanie Birks, Ysanne Chapman, and Karen Francis. 2008. Memoing in qualitative research: Probing data and processes. *Journal of research in nursing* 13, 1 (2008), 68–75.

[7] Georges-Pierre Bonneau, Hans-Christian Hege, Chris R Johnson, Manuel M Oliveira, Kristin Potter, Penny Rheingans, and Thomas Schultz. 2014. Overview and state-of-the-art of uncertainty visualization. *Scientific visualization: Uncertainty, multifield, biomedical, and scalable visualization* (2014), 3–27.

[8] Ali Borji. 2023. Qualitative failures of image generation models and their application in detecting deepfakes. *Image and Vision Computing* 137 (2023), 104771.

[9] Acey Boyce and Tiffany Barnes. 2010. BeadLoom Game: using game elements to increase motivation and learning. In *Proceedings of the fifth international conference on the foundations of digital games*. 25–31.

[10] Michelle Carney, Barron Webster, Irene Alvarado, Kyle Phillips, Noura Howell, Jordan Griffith, Jonas Jongejan, Amit Pitaru, and Alexander Chen. 2020. Teachable machine: Approachable Web-based tool for exploring machine learning classification. In *Extended abstracts of the 2020 CHI conference on human factors in computing systems*. 1–8.

[11] Pew Research Center. 2025. *About a Quarter of US Teens Have Used ChatGPT for Schoolwork, Double the Share in 2023*. https://www.pewresearch.org/short-reads/2025/01/15/about-a-quarter-of-us-teens-have-used-chatgpt-for-schoolwork-double-the-share-in-2023/

[12] Amanda Chaffin and Tiffany Barnes. 2010. Lessons from a course on serious games research and prototyping. In *Proceedings of the Fifth International Conference on the Foundations of Digital Games*. 32–39.

[13] Liuqing Chen, Shuhong Xiao, Yunnong Chen, Yaxuan Song, Ruoyu Wu, and Lingyun Sun. 2024. ChatScratch: an AI-Augmented System Toward Autonomous Visual Programming Learning for Children Aged 6-12. In *Proceedings of the 2024 CHI Conference on Human Factors in Computing Systems* (Honolulu, HI, USA) *(CHI '24)*. Association for Computing Machinery, New York, NY, USA, Article 649, 19 pages. doi:10.1145/3613904.3642229

[14] François Chollet. 2019. On the measure of intelligence. *arXiv preprint arXiv:1911.01547* (2019).

[15] Rudrajit Choudhuri, Ambareesh Ramakrishnan, Amreeta Chatterjee, Bianca Trinkenreich, Igor Steinmacher, Marco Gerosa, and Anita Sarma. 2024. Insights from the Frontline: GenAI Utilization Among Software Engineering Students. (2024).

[16] Douglas B. Clark, Emily E. Tanner-Smith, and Stephen S. Killingsworth. 2016. Digital Games, Design, and Learning: A Systematic Review and Meta-Analysis. *Review of Educational Research* 86, 1 (2016), 79–122. doi:10.3102/0034654315582065 arXiv:https://doi.org/10.3102/0034654315582065 PMID: 26937054.

[17] Daniel C Cliburn. 2006. The effectiveness of games as assignments in an introductory programming course. In *Proceedings. Frontiers in Education. 36th Annual Conference*. IEEE, 6–10.

[18] Aayushi Dangol, Aaleyah Lewis, Hyewon Suh, Xuesi Hong, Hedda Meadan, James Fogarty, and Julie A Kientz. 2025. "I Want to Think Like an SLP": A Design Exploration of AI-Supported Home Practice in Speech Therapy. In *Proceedings of the 2025 CHI Conference on Human Factors in Computing Systems*. 1–21.

[19] Aayushi Dangol, Michele Newman, Robert Wolfe, Jin Ha Lee, Julie A. Kientz, Jason Yip, and Caroline Pitt. 2024. Mediating Culture: Cultivating Socio-cultural Understanding of AI in Children through Participatory Design. In *Proceedings of the 2024 ACM Designing Interactive Systems Conference* (Copenhagen, Denmark) *(DIS '24)*. Association for Computing Machinery, New York, NY, USA, 1805–1822.

[20] Aayushi Dangol, Robert Wolfe, Runhua Zhao, JaeWon Kim, Trushaa Ramanan, Katie Davis, and Julie A. Kientz. 2025. Children's Mental Models of AI Reasoning: Implications for AI Literacy Education. In *Proceedings of the Interaction Design and Children Conference (IDC '25)*. Association for Computing Machinery, Reykjavik, Iceland, 18. doi:10.1145/3713043.3728856

[21] Riddhi Divanji, Aayushi Dangol, Ella J Lombard, Katharine Chen, and Jennifer D Rubin. 2024. Togethertales RPG: Prosocial skill development through digitally mediated collaborative role-playing. In *Proceedings of the 23rd Annual ACM Interaction Design and Children Conference*. 1012–1015.

[22] Stefania Druga. 2018. *Growing up with AI: Cognimates: from coding to teaching machines*. Ph. D. Dissertation. Massachusetts Institute of Technology.

[23] Stefania Druga and Amy J Ko. 2021. How do children's perceptions of machine intelligence change when training and coding smart programs?. In *Proceedings of the 20th annual ACM interaction design and children conference*. 49–61.

[24] Allison Druin. 1999. Cooperative inquiry: developing new technologies for children with children. In *Proceedings of the SIGCHI Conference on Human Factors in Computing Systems* (Pittsburgh, Pennsylvania, USA) *(CHI '99)*. Association for Computing Machinery, New York, NY, USA, 592–599. doi:10.1145/302979.303166

[25] Allison DRUIN. 2002. The role of children in the design of new technology. *Behaviour & information technology* 21, 1 (2002), 1–25.

[26] Utkarsh Dwivedi, Jaina Gandhi, Raj Parikh, Merijke Coenraad, Elizabeth Bonsignore, and Hernisa Kacorri. 2021. Exploring Machine Teaching with Children. In *2021 IEEE Symposium on Visual Languages and Human-Centric Computing (VL/HCC)*, Vol. 2021. IEEE, United States, 1–11.

[27] Jennifer Fereday and Eimear Muir-Cochrane. 2006. Demonstrating rigor using thematic analysis: A hybrid approach of inductive and deductive coding and theme development. *International journal of qualitative methods* 5, 1 (2006), 80–92.

[28] Douglas A Gentile. 2009. Video games affect the brain—for better and worse. *Cerebrum: The DANA Foundation* (2009).

[29] MARK GRIFFITHS. 1997. Computer Game Playing in Early Adolescence. *Youth & society* 29, 2 (1997), 223–237.

[30] Jessica Grose. 2024. What Teachers Told Me About A.I. in School. *The New York Times* (14 Aug. 2024). https://www.nytimes.com/2024/08/14/opinion/ai-schools-teachers-students.html Opinion.

[31] Shuchi Grover. 2024. Teaching AI to K-12 Learners: Lessons, Issues, and Guidance. In *Proceedings of the 55th ACM Technical Symposium on Computer Science Education V. 1* (Portland, OR, USA) *(SIGCSE 2024)*. Association for Computing Machinery, New York, NY, USA, 422–428. doi:10.1145/3626252.3630937

[32] Ariel Han and Zhenyao Cai. 2023. Design implications of generative AI systems for visual storytelling for young learners. In *Proceedings of the 22nd Annual ACM Interaction Design and Children Conference*. 470–474.

[33] Sungsoo Ray Hong, Jessica Hullman, and Enrico Bertini. 2020. Human factors in model interpretability: Industry practices, challenges, and needs. *Proceedings of the ACM on Human-Computer Interaction* 4, CSCW1 (2020), 1–26.

[34] Nicola Jones. 2025. How should we test AI for human-level intelligence? OpenAI's o3 electrifies quest. *Nature* 637, 8047 (2025), 774–775.

[35] Brian Jordan, Nisha Devasia, Jenna Hong, Randi Williams, and Cynthia Breazeal. 2021. PoseBlocks: A toolkit for creating (and dancing) with AI. In *Proceedings of the AAAI Conference on Artificial Intelligence*, Vol. 35. 15551–15559.

[36] Ken Kahn and Niall Winters. 2021. Constructionism and AI: A history and possible futures. *British Journal of Educational Technology* 52, 3 (2021), 1130–1142.

[37] Jeongah Kim and Jaekwoun Shim. 2022. Development of an AR-based AI education app for non-majors. *IEEE Access* 10 (2022), 14149–14156.

[38] Paul A. Kirschner, John Sweller, and Richard E. Clark. 2006. Why Minimal Guidance During Instruction Does Not Work: An Analysis of the Failure of Constructivist, Discovery, Problem-Based, Experiential, and Inquiry-Based Teaching. *Educational psychologist* 41, 2 (2006), 75–86.

[39] Marta Laupa. 1991. Children's Reasoning About Three Authority Attributes: Adult Status, Knowledge, and Social Position. *Developmental psychology* 27, 2 (1991), 321–329.

[40] Michael J Lee. 2014. Gidget: An online debugging game for learning and engagement in computing education. In *2014 ieee symposium on visual languages and human-centric computing (vl/hcc)*. IEEE, 193–194.

[41] Duri Long and Brian Magerko. 2020. What is AI Literacy? Competencies and Design Considerations. In *Proceedings of the 2020 CHI Conference on Human Factors in Computing Systems* (Honolulu, HI, USA) *(CHI '20)*. Association for Computing Machinery, New York, NY, USA, 1–16. doi:10.1145/3313831.3376727

[42] Duri Long, Sophie Rollins, Jasmin Ali-Diaz, Katherine Hancock, Samnang Nuonsinoeun, Jessica Roberts, and Brian Magerko. 2023. Fostering AI Literacy with Embodiment & Creativity: From Activity Boxes to Museum Exhibits. In *Proceedings of the 22nd Annual ACM Interaction Design and Children Conference* (Chicago, IL, USA) *(IDC '23)*. Association for Computing Machinery, New York, NY, USA, 727–731. doi:10.1145/3585088.3594495

[43] Rebecca Marrone, Victoria Taddeo, and Gillian Hill. 2022. Creativity and artificial intelligence—A student perspective. *Journal of Intelligence* 10, 3 (2022), 65.

[44] RE Mayer. 2005. *Cognitive theory of multimedia learning*. The Cambridge Handbook of Visuospatial Thinking/Cambridge University Press.

[45] Richard E. Mayer. 2009. Multimedia learning.

[46] Richard E Mayer and Roxana Moreno. 2003. Nine ways to reduce cognitive load in multimedia learning. *Educational psychologist* 38, 1 (2003), 43–52.

[47] Nora McDonald, Sarita Schoenebeck, and Andrea Forte. 2019. Reliability and Inter-rater Reliability in Qualitative Research: Norms and Guidelines for CSCW and HCI Practice. *Proc. ACM Hum.-Comput. Interact.* 3, CSCW, Article 72 (Nov. 2019), 23 pages. doi:10.1145/3359174

[48] Common Sense Media. 2025. *New Report Shows Students Are Embracing Artificial Intelligence Despite Lack of Parent Awareness*. https://www.commonsensemedia.org/press-releases/new-report-shows-students-are-embracing-artificial-intelligence-despite-lack-of-parent-awareness-and Accessed: 2025-01-26.

[49] Pekka Mertala, Janne Fagerlund, and Oscar Calderon. 2022. Finnish 5th and 6th Grade Students' Pre-Instructional Conceptions of Artificial Intelligence (AI) and Their Implications for AI Literacy Education. 3 (2022), 100095. doi:10.1016/j.caeai.2022.100095

[50] Tilman Michaeli, Stefan Seegerer, Lennard Kerber, and Ralf Romeike. 2023. Data, Trees, and Forests–Decision Tree Learning in K-12 Education. *arXiv preprint arXiv:2305.06442* (2023).

[51] Scott Monteith, Tasha Glenn, John R. Geddes, Peter C. Whybrow, Eric Achtyes, and Michael Bauer. 2024. Artificial intelligence and increasing misinformation. *British journal of psychiatry* 224, 2 (2024), 33–35.

[52] Luis Morales-Navarro, Phillip Gao, Eric Yang, and Yasmin B Kafai. 2024. " It's smart and it's stupid:" Youth's conflicting perspectives on LLMs' language comprehension and ethics. In *Proceedings of the 19th WiPSCE Conference on Primary and Secondary Computing Education Research*. 1–2.

[53] Terran Mott, Alexandra Bejarano, and Tom Williams. 2022. Robot co-design can help us engage child stakeholders in ethical reflection. In *2022 17th ACM/IEEE International Conference on Human-Robot Interaction (HRI)*. IEEE, 14–23.

[54] Michele Newman, Kaiwen Sun, Ilena B Dalla Gasperina, Grace Y Shin, Matthew Kyle Pedraja, Ritesh Kanchi, Maia B Song, Rannie Li, Jin Ha Lee, and Jason Yip. 2024. " I want it to talk like Darth Vader": Helping Children Construct Creative Self-Efficacy with Generative AI. In *Proceedings of the CHI Conference on Human Factors in Computing Systems*. 1–18.

[55] Davy Tsz Kit Ng, Chen Xinyu, Jac Ka Lok Leung, and Samuel Kai Wah Chu. 2024. Fostering students' AI literacy development through educational games: AI knowledge, affective and cognitive engagement. *Journal of computer assisted learning* 40, 5 (2024), 2049–2064.

[56] Mohammad Obaid, Gökçe Elif Baykal, Güncel Kırlangıç, Tilbe Göksun, and Asım Evren Yantaç. 2024. Collective co-design activities with children for designing classroom robots. In *Proceedings of the 4th African Human Computer Interaction Conference* (East London, South Africa) *(AfriCHI '23)*. Association for Computing Machinery, New York, NY, USA, 229–237. doi:10.1145/3628096.3630094

[57] OpenAI. 2024. Early access for safety testing. *OpenAI Blog* , (Dec 2024), .

[58] Allan Paivio and Kalman Csapo. 1973. Picture superiority in free recall: Imagery or dual coding? *Cognitive psychology* 5, 2 (1973), 176–206.

[59] William Christopher Payne, Yoav Bergner, Mary Etta West, Carlie Charp, R. Benjamin Shapiro, Danielle Albers Szafir, Edd V. Taylor, and Kayla DesPortes. 2021. danceON: Culturally Responsive Creative Computing. In *Proceedings of the 2021 CHI Conference on Human Factors in Computing Systems* (Yokohama, Japan) *(CHI '21)*. Association for Computing Machinery, New York, NY, USA, Article 96, 16 pages. doi:10.1145/3411764.3445149

[60] Kristin Potter, Paul Rosen, and Chris R Johnson. 2012. From quantification to visualization: A taxonomy of uncertainty visualization approaches. In *Uncertainty Quantification in Scientific Computing: 10th IFIP WG 2.5 Working Conference, WoCoUQ 2011, Boulder, CO, USA, August 1-4, 2011, Revised Selected Papers*. Springer, 226–249.

[61] Junaid Qadir. 2023. Engineering education in the era of ChatGPT: Promise and pitfalls of generative AI for education. In *2023 IEEE Global Engineering Education Conference (EDUCON)*. IEEE, 1–9.

[62] Kai Quander, Tanzila Roushan Milky, Natalie Aponte, Natalia Caceres Carrascal, and Julia Woodward. 2024. "Are you smart?": Children's Understanding of "Smart" Technologies. In *Proceedings of the 23rd Annual ACM Interaction Design and Children Conference*. 625–638.

[63] Muhammad Raees, Inge Meijerink, Ioanna Lykourentzou, Vassilis-Javed Khan, and Konstantinos Papangelis. 2024. From explainable to interactive AI: A literature review on current trends in human-AI interaction. *International Journal of Human-Computer Studies* (2024), 103301.

[64] Mitchel Resnick. 2024. Generative AI and creative learning: Concerns, opportunities, and choices. (2024).

[65] Mitchel Resnick and Brian Silverman. 2005. Some reflections on designing construction kits for kids. In *Proceedings of the 2005 Conference on Interaction Design and Children* (Boulder, Colorado) *(IDC '05)*. Association for Computing Machinery, New York, NY, USA, 117–122. doi:10.1145/1109540.1109556

[66] Richard Rogers. 2018. Coding and writing analytic memos on qualitative data: A review of Johnny Saldaña's the coding manual for qualitative researchers. *The*

*Qualitative Report* 23, 4 (2018), 889–892.
[67] Donald A. Schön. 1990. *Educating the reflective practitioner: toward a new design for teaching and learning in the professions*. Jossey-Bass, San Francisco.
[68] Marita Skjuve, Asbjørn Følstad, and Petter Bae Brandtzaeg. 2023. The user experience of ChatGPT: findings from a questionnaire study of early users. In *Proceedings of the 5th international conference on conversational user interfaces*. 1–10.
[69] Jaemarie Solyst, Amy Ogan, and Jessica Hammer. 2023. Intergenerational Games to Learn About AI and Ethics. In *Proceedings of the 54th ACM Technical Symposium on Computer Science Education V. 2* (Toronto ON, Canada) *(SIGCSE 2023)*. Association for Computing Machinery, New York, NY, USA, 1273. doi:10.1145/3545947.3573256
[70] Jiahong Su and Weipeng Yang. 2023. Unlocking the power of ChatGPT: A framework for applying generative AI in education. *ECNU Review of Education* 6, 3 (2023), 355–366.
[71] Hyewon Suh, Aayushi Dangol, Hedda Meadan, Carol A Miller, and Julie A Kientz. 2024. Opportunities and challenges for AI-based support for speech-language pathologists. In *Proceedings of the 3rd Annual Meeting of the Symposium on Human-Computer Interaction for Work*. 1–14.
[72] Yujie Sun, Dongfang Sheng, Zihan Zhou, and Yifei Wu. 2024. AI hallucination: towards a comprehensive classification of distorted information in artificial intelligence-generated content. *Humanities and Social Sciences Communications* 11, 1 (2024), 1–14.
[73] John Sweller. 2010. Element Interactivity and Intrinsic, Extraneous, and Germane Cognitive Load. *Educational psychology review* 22, 2 (2010), 123–138.
[74] Lev Tankelevitch, Viktor Kewenig, Auste Simkute, Ava Elizabeth Scott, Advait Sarkar, Abigail Sellen, and Sean Rintel. 2024. The metacognitive demands and opportunities of generative AI. In *Proceedings of the 2024 CHI Conference on Human Factors in Computing Systems*. 1–24.
[75] Anastasios Theodoropoulos. 2022. Participatory design and participatory debugging: Listening to students to improve computational thinking by creating games. *International Journal of Child-Computer Interaction* 34 (2022), 100525.
[76] David Touretzky, Christina Gardner-McCune, Fred Martin, and Deborah Seehorn. 2019. Envisioning AI for K-12: What Should Every Child Know about AI? *Proceedings of the AAAI Conference on Artificial Intelligence* 33, 01 (Jul. 2019), 9795–9799. doi:10.1609/aaai.v33i01.33019795
[77] Jessica Van Brummelen, Viktoriya Tabunshchyk, and Tommy Heng. 2021. "Alexa, Can I Program You?": Student Perceptions of Conversational Artificial Intelligence Before and After Programming Alexa. In *Proceedings of the 20th Annual ACM Interaction Design and Children Conference* (Athens, Greece) *(IDC '21)*. Association for Computing Machinery, New York, NY, USA, 305–313. doi:10.1145/3459990.3460730
[78] Iro Voulgari, Marvin Zammit, Elias Stouraitis, Antonios Liapis, and Georgios Yannakakis. 2021. Learn to Machine Learn: Designing a Game Based Approach for Teaching Machine Learning to Primary and Secondary Education Students. In *Proceedings of the 20th Annual ACM Interaction Design and Children Conference* (Athens, Greece) *(IDC '21)*. Association for Computing Machinery, New York, NY, USA, 593–598. doi:10.1145/3459990.3465176
[79] Lev S Vygotsky. 1978. Mind in society (M. Cole, V. John-Steiner, S. Scribner, & E. Souberman, Eds.).
[80] Greg Walsh, Elizabeth Foss, Jason Yip, and Allison Druin. 2013. FACIT PD: a framework for analysis and creation of intergenerational techniques for participatory design. In *proceedings of the SIGCHI Conference on Human Factors in Computing Systems*. 2893–2902.
[81] Dakuo Wang, Elizabeth Churchill, Pattie Maes, Xiangmin Fan, Ben Shneiderman, Yuanchun Shi, and Qianying Wang. 2020. From human-human collaboration to Human-AI collaboration: Designing AI systems that can work together with people. In *Extended abstracts of the 2020 CHI conference on human factors in computing systems*. 1–6.
[82] Heidi C Webb and Mary Beth Rosson. 2011. Exploring careers while learning Alice 3D: a summer camp for middle school girls. In *Proceedings of the 42nd ACM technical symposium on Computer science education*. 377–382.
[83] Merlin C. Wittrock. 1989. Generative Processes of Comprehension. *Educational psychologist* 24, 4 (1989), 345–376.
[84] Robert Wolfe, Aayushi Dangol, Alexis Hiniker, and Bill Howe. 2024. Dataset Scale and Societal Consistency Mediate Facial Impression Bias in Vision-Language AI. In *Proceedings of the AAAI/ACM Conference on AI, Ethics, and Society*, Vol. 7. 1635–1647.
[85] Robert Wolfe, Aayushi Dangol, Bill Howe, and Alexis Hiniker. 2024. Representation Bias of Adolescents in AI: A Bilingual, Bicultural Study. In *Proceedings of the AAAI/ACM Conference on AI, Ethics, and Society*, Vol. 7. 1621–1634.
[86] Robert Wolfe and Tanushree Mitra. 2024. The Impact and Opportunities of Generative AI in Fact-Checking. In *The 2024 ACM Conference on Fairness, Accountability, and Transparency*. 1531–1543.
[87] Julia Woodward, Feben Alemu, Natalia E. López Adames, Lisa Anthony, Jason C. Yip, and Jaime Ruiz. 2022. "It Would Be Cool to Get Stampeded by Dinosaurs": Analyzing Children's Conceptual Model of AR Headsets Through Co-Design. In *Proceedings of the 2022 CHI Conference on Human Factors in Computing Systems*. 1–13.
[88] Julia Woodward, Zari McFadden, Nicole Shiver, Amir Ben-Hayon, Jason C Yip, and Lisa Anthony. 2018. Using co-design to examine how children conceptualize intelligent interfaces. In *Proceedings of the 2018 CHI conference on human factors in computing systems*. 1–14.
[89] Yi Wu. 2023. Integrating generative AI in education: how ChatGPT brings challenges for future learning and teaching. *Journal of Advanced Research in Education* 2, 4 (2023), 6–10.
[90] Lixiang Yan, Samuel Greiff, Ziwen Teuber, and Dragan Gašević. 2024. Promises and challenges of generative artificial intelligence for human learning. *Nature Human Behaviour* 8, 10 (2024), 1839–1850.
[91] Robert K Yin. 2013. Validity and generalization in future case study evaluations. *Evaluation* 19, 3 (2013), 321–332.
[92] Jason C Yip, Frances Marie Tabio Ello, Fumi Tsukiyama, Atharv Wairagade, and June Ahn. 2023. " Money shouldn't be money!": An Examination of Financial Literacy and Technology for Children Through Co-Design. In *Proceedings of the 22nd Annual ACM Interaction Design and Children Conference*. 82–93.
[93] Jason C. Yip, Kiley Sobel, Caroline Pitt, Kung Jin Lee, Sijin Chen, Kari Nasu, and Laura R. Pina. 2017. Examining Adult-Child Interactions in Intergenerational Participatory Design. In *Proceedings of the 2017 CHI Conference on Human Factors in Computing Systems* (Denver, Colorado, USA) *(CHI '17)*. Association for Computing Machinery, New York, NY, USA, 5742–5754. doi:10.1145/3025453.3025787
[94] Eunice Yiu, Eliza Kosoy, and Alison Gopnik. 2024. Transmission Versus Truth, Imitation Versus Innovation: What Children Can Do That Large Language and Language-and-Vision Models Cannot (Yet). *Perspectives on Psychological Science* 19, 5 (2024), 874–883. doi:10.1177/17456916231201401 arXiv:https://doi.org/10.1177/17456916231201401 PMID: 37883796.
[95] Yen Na Yum, Neil Cohn, and Way Kwok-Wai Lau. 2021. Effects of picture-word integration on reading visual narratives in L1 and L2. *Learning and Instruction* 71 (2021), 101397.
[96] Chao Zhang, Cheng Yao, Jianhui Liu, Zili Zhou, Weilin Zhang, Lijuan Liu, Fangtian Ying, Yijun Zhao, and Guanyun Wang. 2021. StoryDrawer: A Co-Creative Agent Supporting Children's Storytelling through Collaborative Drawing. In *Extended Abstracts of the 2021 CHI Conference on Human Factors in Computing Systems*. 1–6.
[97] Jiahua Zhao, Gwo-Jen Hwang, Shao-Chen Chang, Qi-fan Yang, and Artorn Nokkaew. 2021. Effects of gamified interactive e-books on students' flipped learning performance, motivation, and meta-cognition tendency in a mathematics course. *Educational Technology Research and Development* 69 (2021), 3255–3280.
[98] Christopher Zorn, Chadwick A Wingrave, Emiko Charbonneau, and Joseph J LaViola Jr. 2013. Exploring Minecraft as a conduit for increasing interest in programming.. In *FDG*. 352–359.